\documentclass[12pt,a4paper,final]{iopart}

\usepackage{iopams} 
\expandafter\let\csname equation*\endcsname\relax
\expandafter\let\csname endequation*\endcsname\relax
\usepackage{amsmath, amssymb}
\usepackage{graphicx}
\usepackage{subfig}
\usepackage[breaklinks=true,colorlinks=true,linkcolor=blue,urlcolor=blue,citecolor=blue]{hyperref}
\usepackage{color}

\usepackage[square,sort&compress,numbers]{natbib}
\usepackage{braket}
\usepackage{placeins}

\usepackage[latin1]{inputenc}

\newcommand{\bit}{\begin{itemize} \setlength{\itemsep}{0ex} \setlength{\topsep}{0ex} } 
\newcommand{\eit}{\end{itemize}}
\newcommand{\be}{\begin{equation}}
\newcommand{\ee}{\end{equation}}
\newcommand{\bea}{\begin{eqnarray}}
\newcommand{\eea}{\end{eqnarray}}
\newcommand{\ba}{\begin{align}}
\newcommand{\ea}{\end{align}}
\newcommand{\SKIP}[1]{}

\def \ra{\rightarrow}
\def \ua{\uparrow}
\def \da{\downarrow}
\def \la{\leftarrow}

\begin{document}

\title[Interaction effects in spin-orbit wires]{Interaction effects in a microscopic quantum wire model with strong spin-orbit interaction}

\author{G W Winkler$^{1,2}$, M Ganahl$^{3,2}$, D Schuricht$^4$, H G Evertz$^{2,5}$ and S Andergassen$^{6,7}$}
\address{$^1$ Theoretical Physics and Station Q Zurich, ETH Zurich, 8093 Zurich, Switzerland}
\address{$^2$ Institute of Theoretical and Computational Physics, Graz University of Technology, 8010 Graz, Austria}
\address{$^3$ Perimeter Institute for Theoretical Physics, Waterloo, Ontario N2L 2Y5, Canada}
\address{$^4$ Institute for Theoretical Physics, Center for Extreme Matter and Emergent Phenomena, Utrecht University, Princetonplein 5, 3584 CE Utrecht, The Netherlands}
\address{$^5$ Kavli Institute for Theoretical Physics, University of California, Santa Barbara, CA 93106, USA}
\address{$^6$ Institute for Theoretical Physics and Center for
  Quantum Science, Universit\"at T\"ubingen, Auf der Morgenstelle 14, D-72076 T\"ubingen, Germany}
\address{$^7$ Faculty of Physics, University of Vienna, Boltzmanngasse 5,  1090 Vienna, Austria}
\ead{winklerg@phys.ethz.ch}

\date{\today}

\begin{abstract}
  We investigate the effect of strong interactions on the spectral properties
  of quantum wires with strong Rashba spin-orbit interaction in a magnetic field,
  using a combination of Matrix Product State and bosonization techniques. Quantum
  wires with strong Rashba spin-orbit interaction and magnetic field exhibit a partial
  gap in one-half of the conducting modes. Such systems have attracted wide-spread
  experimental and theoretical attention due to their unusual physical properties, among
  which are spin-dependent transport, or a topological superconducting phase
  when under the proximity effect of an $s$-wave superconductor. As a microscopic model
  for the quantum wire we study an extended Hubbard model with spin-orbit interaction
  and Zeeman field. We obtain spin resolved spectral densities from the real-time
  evolution of excitations, and calculate the phase diagram.
  We find that interactions increase the pseudo gap at $k = 0$ and thus also enhance
  the Majorana-supporting phase and stabilize the helical spin order. Furthermore, we
  calculate the optical conductivity and compare it with the low energy spiral Luttinger
  Liquid result, obtained from field theoretical calculations. With interactions, the optical
  conductivity is dominated by an excotic excitation of a bound soliton-antisoliton pair
  known as a breather state. We visualize the oscillating motion of the breather state,
  which could provide the route to their experimental detection in e.g. cold atom
  experiments.
\end{abstract}

\pacs{71.10.Pm,71.10.-w,71.35.Cc,71.70.Ej}

\maketitle


\eqnobysec

\section{\label{sec:level1}Introduction}

Electron-electron (e-e) interactions have drastic effects on the
physics of 1D electron conductors.  The low energy properties of
interacting 1D systems are described by the Luttinger liquid (LL)
model which gives a qualitative picture of the static and dynamic
properties~\cite{giamarchi2004quantum, Voit_One-dimensional_1995,
  Meden_Spectral_1992, Schoenhammer_Luttinger_2002}.  The LL
parameters $K_\rho$, $K_\sigma$, $v_\rho$ and $v_\sigma$ for a given
microscopic model are, however, not easy to obtain in general.
Furthermore, simplifications like the linear Dirac dispersion impede
the application of LL theory away from the low energy
regime\footnote{There are some extensions trying to overcome this
  limitation~\cite{Imambekov_Universal_2009,imambekov_ll_2012}.}.

In the present paper we use density matrix renormalization group
(DMRG)\cite{white_density_1992, white_density_1993} techniques to
investigate the combined effect of Rashba spin-orbit (SO)
interaction~\cite{SO_review} and Zeeman field on one-dim. interacting
quantum wires in a microscopic model, at all energies. It is known
that the combination of SO and Zeeman field in a 1D quantum wire, with
superconductivity induced by the proximity effect, supports Majorana
zero modes --- elusive quasiparticles which are their own
antiparticles --- at the boundaries of the
system~\cite{oreg_helical_2010,lutchyn_majorana_2010}. Due to their
non-Abelian exchange statistics Majorana zero modes are promising
candidates for the implementation of topologically protected quantum
information processing~\cite{top_comp1,top_comp2}. Considerable
experimental effort is currently under way to investigate the physics
of Majorana zero modes in 1D semiconductor nanowires with strong
spin-orbit and Zeeman
fields~\cite{das_majorana_2012,xu_majorana_2012,mourik_majorana_2012,CM_Majorana}.

In the non-interacting case, the SO interaction generates spin-split
bands, with crossing points of two bands of orthogonal spin at $k=0$.
A Zeeman field then lifts the degeneracy at the crossing points,
leading to the formation of a spin-orbit (SO) gap.  Experiments in
high-mobility GaAs/AlGaAs hole and InAs electron quantum wires have
observed this gap~\cite{quay_observation_2010,Heedt_helical_gap_2017}.
Apart from topological quantum computation, these properties make such
systems also ideal candidates for spintronic
devices~\cite{birkholz2008spin} like spin
filters~\cite{streda_antisymmetric_2003,Mazza_Spin_2013} or Cooper
pair splitters~\cite{sato_cooper_2010}.  Furthermore, the effects of
strong correlations in spin-orbit chains can be investigated in
artificially engineered cold atoms
chains~\cite{cheuk_coldatoms_2012,wang_coldatoms_2012}.

One of the goals of this work is to compare the properties of our
microscopic model to those of the popular LL
models~\cite{Schulz_Low-energy_2009,PhysRevB.80.041308,PhysRevB.82.033407,PhysRevB.84.075466,2016arXiv160808143T,PhysRevB.94.125420,PhysRevB.94.245414,PhysRevB.66.073311,PhysRevB.91.235421,gangadharaiah,Governale2004581,PhysRevLett.94.137207},
specifically, the spiral LL and the helical LL. The spiral LL was
first introduced by Braunecker
et~al.~\cite{braunecker_nuclear_2009,braunecker_nuclear_2009_1} to
describe a LL embedded in a lattice of nuclear spins.  The hyperfine
interactions between the nuclear and electron spins trigger a strong
feedback reaction that gives an effective helical magnetic field
spiraling along the wire which opens a partial gap like the SO gap.
Later it was realized~\cite{braunecker_spin-selective_2010} that the
same low energy model applies to systems with SO interaction and
Zeeman field if the momentum shift of the SO interaction is
commensurable with the Fermi momentum: $k_\text F=2k_\text{SO}$.
Indeed, the Hamiltonians underlying these two systems are linked by a
simple gauge transformation~\cite{braunecker_spin-selective_2010}.
The spectral properties of the spiral LL were further
investigated~\cite{braunecker_spectral_2012,schuricht_spectral_2011}
and it was shown that they can be approximated by a simpler helical LL
for energies much smaller than the gap. In this limit the system thus
becomes effectively spinless, which is reflected by our
results. Furthermore, the transport properties of the spiral LL with
and without attached leads have been
studied~\cite{meng_low_2014}. Experimental evidence for the spiral LL
has been given in Ref.~\cite{scheller_spiral_experiment}.

The paper is organized as follows.  In section \ref{sec2} we introduce
the microscopic model and describe the method used to determine the
spectral functions. Results are discussed in section \ref{sec3}.  In
particular, we present a detailed analysis of the phase diagram and
show that the helical phase is stabilized by the interactions.  We
also show results for the charge, spin and current structure factors
as well as for the optical conductivity, which exhibits peculiar
breather modes.  Based on our numerical study we compare the results
to field-theoretical predictions and discuss the applicability of the
LL description.  We conclude with a summary and an outlook in section
\ref{sec4}.

\section{\label{sec:level2}Model and Methods}
\label{sec2}

\subsection{Microscopic model}
In this work we consider a one-dimensional quantum wire, with strong
Rashba spin orbit coupling in a magnetic field. In addition, we are
interested in the effects of strong electronic interactions, which we
model using an interaction of the extended-Hubbard type. The
Hamiltonian for this system is then given by
\begin{equation}
  \begin{aligned}
    H_\mathrm{t}   &= \sum_j \left[-\frac{t}{2}\, \left(c_j^{\dag}c_{j+1}+h.c.\right)-\left(\mu-t\right) \, c_j^{\dag}c_j\right] ,  \\
    H_\mathrm{SO}  &= \sum_j \left[ -\frac{\alpha}{2} \,\left(ic_j^{\dag}\sigma^y c_{j+1}+h.c.\right) \right] , \\
    H_\mathrm{B}   &= \sum_j B \, c_j^{\dag} \sigma^z c_j , \\
    H_\mathrm{int} &= \sum_j \left( U \,\hat n_{j\uparrow}\hat
      n_{j\downarrow}+U' \, \hat n_j\hat n_{j+1}\right) ,\;
  \end{aligned}
  \label{eq:hamiltonian}
\end{equation}
where $c^\dag_j=(c^\dag_{j\ua},c^\dag_{j\da})$ are spinors and
$\sigma^y$ as well as $\sigma^z$ are the Pauli matrices acting in
spinor space. Here, $H_\mathrm{t}$ corresponds to the kinetic and
potential energy, $H_\mathrm{SO}$ to the Rashba SO coupling, and
$H_\mathrm{B}$ to the magnetic field. $H_\mathrm{int}$ contains local
and nearest-neighbor interaction $U$ and $U'$, respectively, with the
densities $\hat n_{j\sigma}=c_{j\sigma}^\dagger c_{j\sigma}$ and
$\hat n_j=\hat n_{j\uparrow}+\hat n_{j\downarrow}$.  The Zeeman field
$B$ is applied in $z$ direction, and we choose the quantization axis
of the SO coupling in the $y$ direction. In the present paper all
parameters are constant throughout the wire, the inclusion of
disorder, to account for a more realistic model of a quantum
wire~\cite{Sherman1,Sherman2}, will be part of future work. Throughout
the paper we will use $t=1$ as the unit of energy. Our model can also
be seen as a low energy description of the conduction band in a
semiconductor nanowire, discretized on a coarser lattice than the
actual atomic positions. Via rescaling the energies and discretization
length scale, our results can be directly applied for a wide range of
experimental parameters~\cite{parameters}.

\begin{figure}[t]
  \center
  \subfloat[Dispersion]{\includegraphics[width=0.5\textwidth]{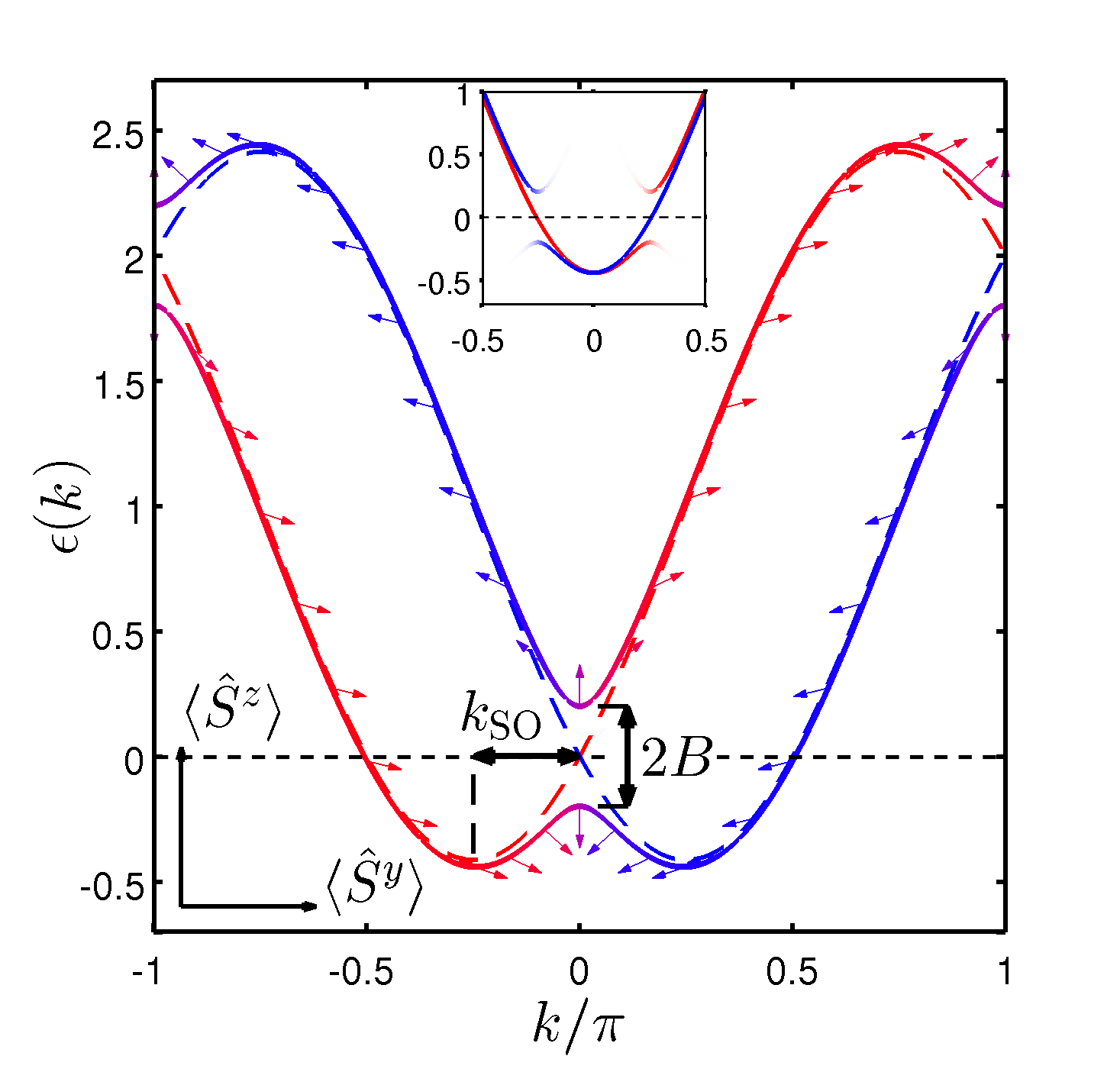}}
  {\small \raisebox{2.7cm}[0pt]{\subfloat[Helical
      LL]{\def\svgwidth{0.24\textwidth}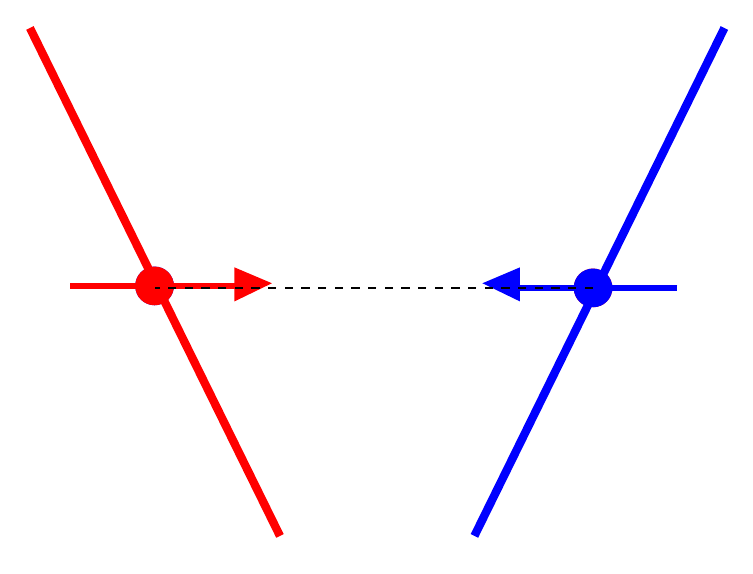}\qquad
      \subfloat[Spiral
      LL]{\def\svgwidth{0.24\textwidth}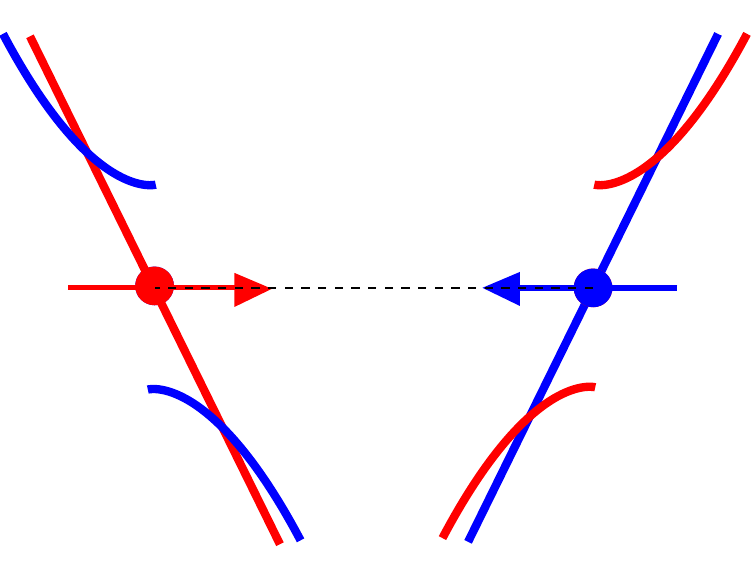}}}
  \caption{(a) Dispersion $\epsilon(k)$ of the non-interacting
    Hamiltonian $H_0$ with $\alpha=1$, for $B=0.2$ (solid line) and
    $B=0$ (dashed line). The arrows indicate the spin direction of the
    eigenvectors (shown only for $B=0.2$), with positive (negative)
    $\langle \hat S^y \rangle$ eigenstates displayed in red (blue).
    The inset shows the dispersion in the gauge transformed basis
    (eq.~\eqref{eq:gauge}).  Figures (b) and (c) show the dispersions
    of the corresponding low-energy LLs. The chemical potential is
    marked by the horizontal dashed lines. }
  \label{fig:disp}
\end{figure}

For the purpose of analyzing the spectral functions for the
interacting model, it is illuminating to first discuss the spectral
properties of the non-interacting part
$H_0=H_\mathrm{t}+H_\mathrm{SO}+H_\mathrm{B}$
\cite{oreg_helical_2010,lutchyn_majorana_2010,stoudenmire_interaction_2011}.
Figure \ref{fig:disp} shows the dispersion of $H_0$ for two different
cases.  The dashed lines show the dispersion with non-zero SO coupling
and no Zeeman field ($B=0$).  Due to the SO coupling, the two Kramers
degenerate bands of $H_\mathrm{t}$ split up into two branches, one for
each eigenstate of $\hat S^y$. The branches are shifted to the left,
respectively right, by a momentum

\begin{equation}\label{eq:kso}
  k_\text{SO}=\arctan\left(\frac \alpha t\right).
\end{equation}
Upon turning on the Zeeman field $B$, a gap opens at the crossing
points of the two branches at $k=0$ and $k=\pi$, due to the now
non-zero coupling between the two branches. The so-called SO gap at
$k=0$ is of size $2B$.  The ground state has either two Fermi points
(2F phase) when the chemical potential is tuned to lie inside the SO
gap, or four Fermi points (4F phases) when the chemical potential lies
below or above the SO gap (and below the second gap at $k=\pi$).  In
the so-called helical 2F phase opposite Fermi points have
approximately orthogonal spin directions. This is the regime we focus
on in this paper. The approximately opposite spin direction at the two
Fermi points has some interesting physical implications for e.g. low
energy electronic transport, where only excitations close to the Fermi
points are relevant. Due to the opposite electron velocities at
opposite Fermi points, the charge transport in this regime is highly
spin dependent.  A right-moving current can only carry negative
$\hat S^y$ electrons, and vice versa for a left-moving current.
Another implication is the appearance of a topological,
superconducting ground state if an $s$-wave pairing term
$c^\dag_\uparrow c^\dag_\downarrow + h.c.$ is added to the
Hamiltonian.  In this case, the system becomes a topological $p$-wave
superconductor, since only electrons at $+k_\mathrm{F}$ and
$-k_\mathrm{F}$ are available for pairing. It has been shown that the
quantum wire in this case hosts Majorana fermions at its
edges~\cite{kitaev,oreg_helical_2010,lutchyn_majorana_2010}.  To
optimally access this phase, the Fermi energy has to be tuned to the
center of the SO gap, which is the case if the Fermi momentum
$k_F=2k_{SO}$.  Like for spinless fermions, the relation between the
average number of electrons per site $n$ and the Fermi momentum is
simply $k_\text F=n\pi$ inside the SO gap. For $\alpha=1$ (see
Eq.(\ref{eq:kso})), we get $k_\text{SO}=\pi/4$ and hence the middle of
the 2F phase is reached at quarter filling $n=0.5$. One goal of this
work is to investigate the effects of electronic correlations on the
stability of this 2F phase (see also
\cite{stoudenmire_interaction_2011,Mazza_Laughlin_2016,Mazza_Synthetic_2016,Mazza_Magnetic_2015}
for related work).

Interactions can be included in several different ways. In this work,
we use time-dependent MPS techniques to compute spectral functions,
and compare our results to analytic approaches using bosonization
techniques. To obtain better comparability to the LL results, we
choose a nearest neighbor interaction $U'=U/2$ in our numerical
calculations (apart from \ref{us0}).  This choice minimizes two
particle backscattering (scattering from $+k_\text F$ to
$-k_\text F$), which is a major source of deviation from LL behavior
in finite size systems.  The two-particle backscattering parameter for
the extended Hubbard model is approximately given by
$g_{1\perp}=U+2U'\cos(2k_\text F)$
\cite{andergassen_renormalization-group_2006}, which vanishes for
$U'=\frac U 2$ for $k_\text F=\pi/2$.  In the most part of the present
work numerical results are obtained for quarter filling, amounting to
$k_\text F=\pi/2$ in the 2F phase.

In the 2F phase, the low energy physics of our model is captured well
by LL theory.  In this respect, there exist two related approaches,
the helical and the spiral LL \cite{braunecker_spectral_2012}.  The
underlying free dispersions are shown in figure \ref{fig:disp} (b) and
(c).  In the case of the helical LL, the similarity to the dispersion
in figure \ref{fig:disp} (a) in the 2F phase is evident.  Figure
\ref{fig:disp} (b) and (c) have orthogonal spin directions at opposite
Fermi points.  The spiral LL figure \ref{fig:disp} (c) additionally
includes gapped modes. The helical LL may be seen as a low energy
model of the spiral LL as long as the relevant excitation energies
$\omega$ are smaller than the gap $2B$.  The helical LL is very
similar to the spinless fermion LL and many properties like
density-density correlations are indeed the same.

The relation between the helical and spiral LL approach can be
clarified in our model by applying a gauge transformation
\cite{braunecker_spin-selective_2010}
\begin{equation}
  c_j\ra e^{-i\sigma_y k_\text{SO} j} c_j\;.
  \label{eq:gauge}
\end{equation}
to our model Hamiltonian Eq.(\ref{eq:hamiltonian}). In the
non-interacting case, this yields a modified dispersion relation, as
depicted in the inset of figure \ref{fig:disp} (a), from which the
relation to the spiral LL becomes apparent.  By applying this
transformation to Eq.~\eqref{eq:hamiltonian}, the SO interaction
vanishes, $\tilde H_\mathrm{SO}=0$, while $H_\text{t}$ and
$H_\text{B}$ in Eq.~\eqref{eq:hamiltonian} become
\begin{align}
  \begin{split}
    \tilde H_\mathrm{t}   &= \sum_j \left[ -\frac{\sqrt{t^2+\alpha^2}}{2} (c_j^{\dag}c_{j+1}+h.c.)-(\mu-t)c_j^{\dag}c_j \right] \\
    \tilde H_\mathrm{B} &= \sum_j B \left[ \cos(2 k_\text{SO} j)
      c_j^{\dag} \sigma^z c_j - \sin(2 k_\text{SO} j) c_j^{\dag}
      \sigma^x c_j \right]\;.
  \end{split}
\end{align}
Note how the site independent magnetic field transforms into a helical
field spiraling around the quantum wire, hence the name spiral LL
~\cite{braunecker_spectral_2012,braunecker_nuclear_2009,braunecker_nuclear_2009_1}.
The kinetic energy gets rescaled with $t\ra
\sqrt{t^2+\alpha^2}$. Since $H_\mathrm{int}$ is unaffected by this
transformation, our results are valid for both cases in the
commensurate case $k_\text F= 2 k_\text{SO}$.

\subsection{Spectral functions from real-time evolution}
In this work we are analyzing the effect of e-e interactions on the
spectral properties of a quantum wire, using Matrix Product State
(MPS) \cite{fannes_finitely_1992,schollwoeck_density-matrix_2010}
techniques.  MPS are nowadays routinely used in calculations of
spectral functions in 1d quantum system. Recent advances include
Chebyshev
\cite{weisse_chebyshef_2006,holzner_chebyshev_2011,thomale_chebyshev_2013,braun_chebyshev_2014,ganahl_chebyshev_2014,PhysRevB.91.115144}
and Lanczos \cite{dargel_lanczos_2012,PhysRevE.94.053308} expansion techniques. Here, we
obtain real-frequency Greens functions using real time evolution
employing the Time Dependent Block Decimation (TEBD)
\cite{vidal_efficient_2003, vidal_efficient_2004} and subsequent
Fourier transformation
\cite{white_spectral_2008,barthel_spectral_2009,Ganahl_Efficient_2014,PhysRevB.90.245127}. The
object of interest is the Greens function
\begin{equation}
  S_{\hat A \hat B}(k,\omega)= \sum_{j=-\infty}^\infty e^{-ikj} \int_{-\infty}^\infty dt\, e^{i\omega t} \left(\langle 0| \frac{1}{2} \{\hat A_j(t), \hat B_0(0)\} |0\rangle-\langle 0|\hat A_j(0)|0\rangle \langle 0|\hat B_0(0) |0\rangle\right),
  \label{eq:skw}
\end{equation}
where $\hat A_j(t)$ and $\hat B_j(t)$ are operators acting on site $j$
at time $t$, $\{\hat A, \hat B\} = \hat A \hat B + \hat B \hat A$ is
the anticommutator and $\ket{0}$ is the ground state.  For our case of
spinors, with $\hat A_j = c_j$ and $\hat B_0 = c^\dag_0$, we obtain
the spectral function $S_{cc^\dag}(k,\omega)$.  The computation
proceeds as follows: first, we compute the ground state $\ket{0}$
using DMRG. $S_{cc^{\dagger}}(k,w)$ is then computed from Fourier
transformation of the time dependent functions
\begin{equation}\label{eq:green}
  G_{i\sigma,j\sigma'}(\omega)\equiv\bra{0}c_{i\sigma}^{\dagger}(t)c_{j\sigma'}(0)\ket{0}=e^{iE_0t}\bra{0}c_{i\sigma}^{\dagger}e^{-iHt}
  c_{j\sigma'}\ket{0}
\end{equation}
which are calculated by evolving
$\ket{\psi(t)}\equiv e^{-iHt}c_{j\sigma'}\ket{0}$ forward in time, and
calculating the overlap with $\bra{0}c_{i\sigma}^{\dagger}$. The phase
factor $e^{iE_0t}$ can be removed by shifting the ground state energy
$E_0$ of $\ket{0}$ to 0 prior to the evolution.

A common feature shared by all MPS methods for calculating spectral
functions is the finite $\omega$-resolution. In our case it is due to
the fact that only short to moderately long time scales can be reached
with MPS time evolution. This is due to rapid entanglement growth
following (local) quenches of the system.  The $\omega$-resolution can
be substantially improved by using extrapolation techniques for the
time series. In this paper we use the so-called linear prediction
technique to achieve
this~\cite{white_spectral_2008,barthel_spectral_2009,Ganahl_Efficient_2014},
see also \ref{linear_prediction}. For our study we use system sizes of
up to $L=256$ sites, and matrix dimension up to $m=600$ and $m=1200$
for DMRG and during time evolution, respectively, which was large
enough to ensure that our results did not depend on the bond dimension
$m$ any more.  In our code we make use of total charge conservation
(note that total $S^z$ is not conserved during the time evolution).
For the time evolution, we use a second order Suzuki-Trotter splitting
scheme, with a time step of $0.05$ ($0.01$ for breather calculations,
see below).  The dominant error is due to the truncation of the MPS
matrices during time evolution. A possible measure of this error is
the cumulative truncated weight \cite{white_spectral_2008}
\begin{equation}
  \epsilon_\text{tot}(t)=\sum_{\tau=0}^{t} \sum_{j=1}^{L} \epsilon_j(\tau)\;,
\end{equation}
where the sums run over lattice sites $j$ and time slices $\tau$.
Note that the growth of entanglement, and hence
$\epsilon_\text{tot}(t)$ depend on the particular kind of quench that
is applied to the system. For example, for single-particle spectral
functions entanglement growth is stronger than for the charge and spin
structure factors or the current-current correlation
function. Furthermore, for strong interactions the entanglement growth
during the time evolution is more pronounced than for weak
interactions, while ground states for stronger interactions typically
require smaller bond dimensions.

An advantage of our method is the direct accessibility of real
frequency spectra without the need of an analytic continuation like in
QMC based approaches \cite{Assaad_Computational_2008}.  It is also
considerably more efficient than previous approaches like Dynamical DMRG
\cite{jeckelmann_density_2002} or the Correction Vector Method
\cite{kuhner_dynamical_1999}.  In our approach, spectra at all momenta
can be calculated from a single time evolution.

\section{\label{sec:level3}Results and Discussion}
\label{sec3}
\subsection{Spectral functions}
\begin{figure}[t]
  \centering
  \subfloat[]{\includegraphics[width=0.66\textwidth]{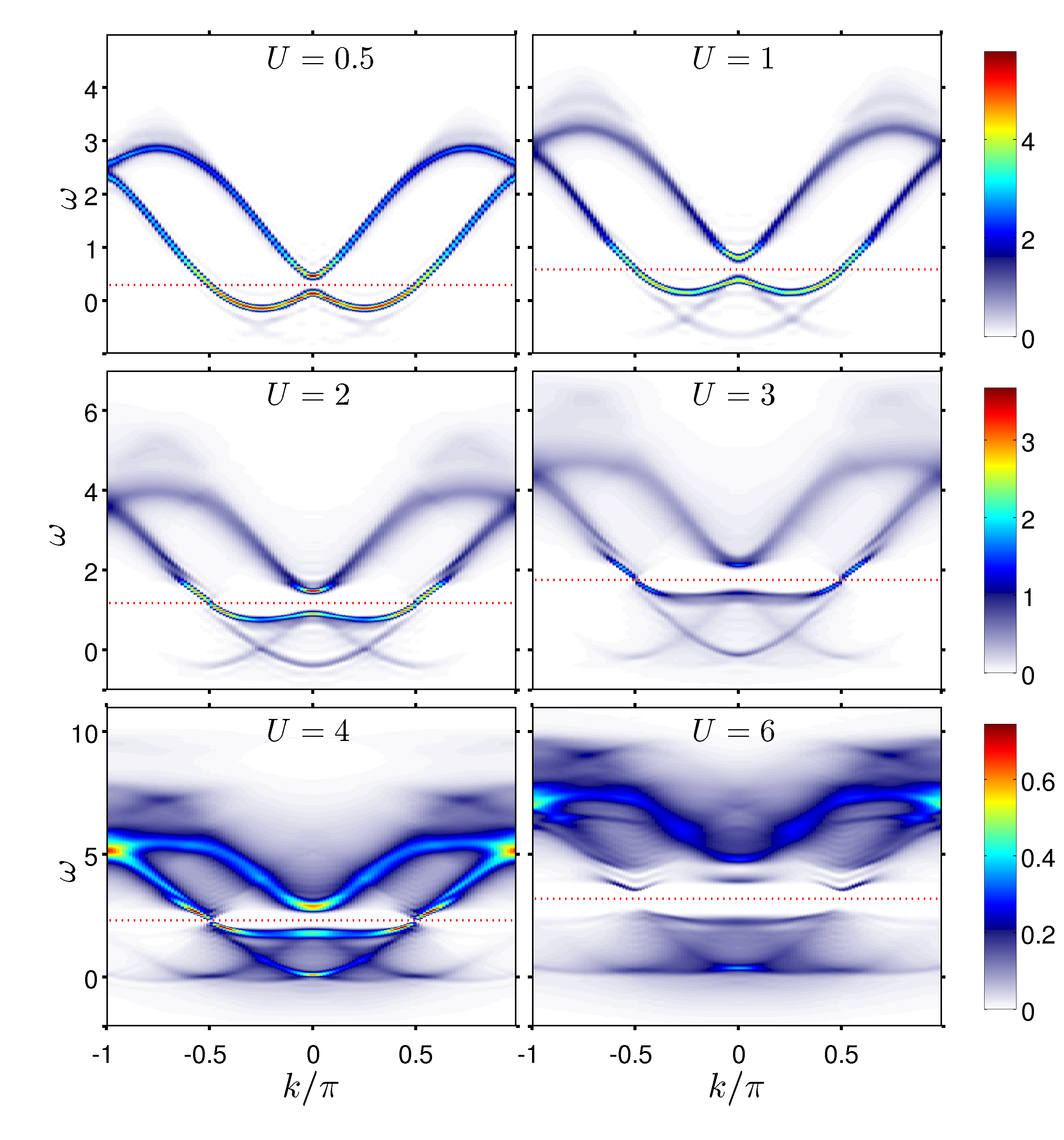}}
  \raisebox{3.8cm}[0pt]{\subfloat[]{\includegraphics[width=0.33\textwidth]{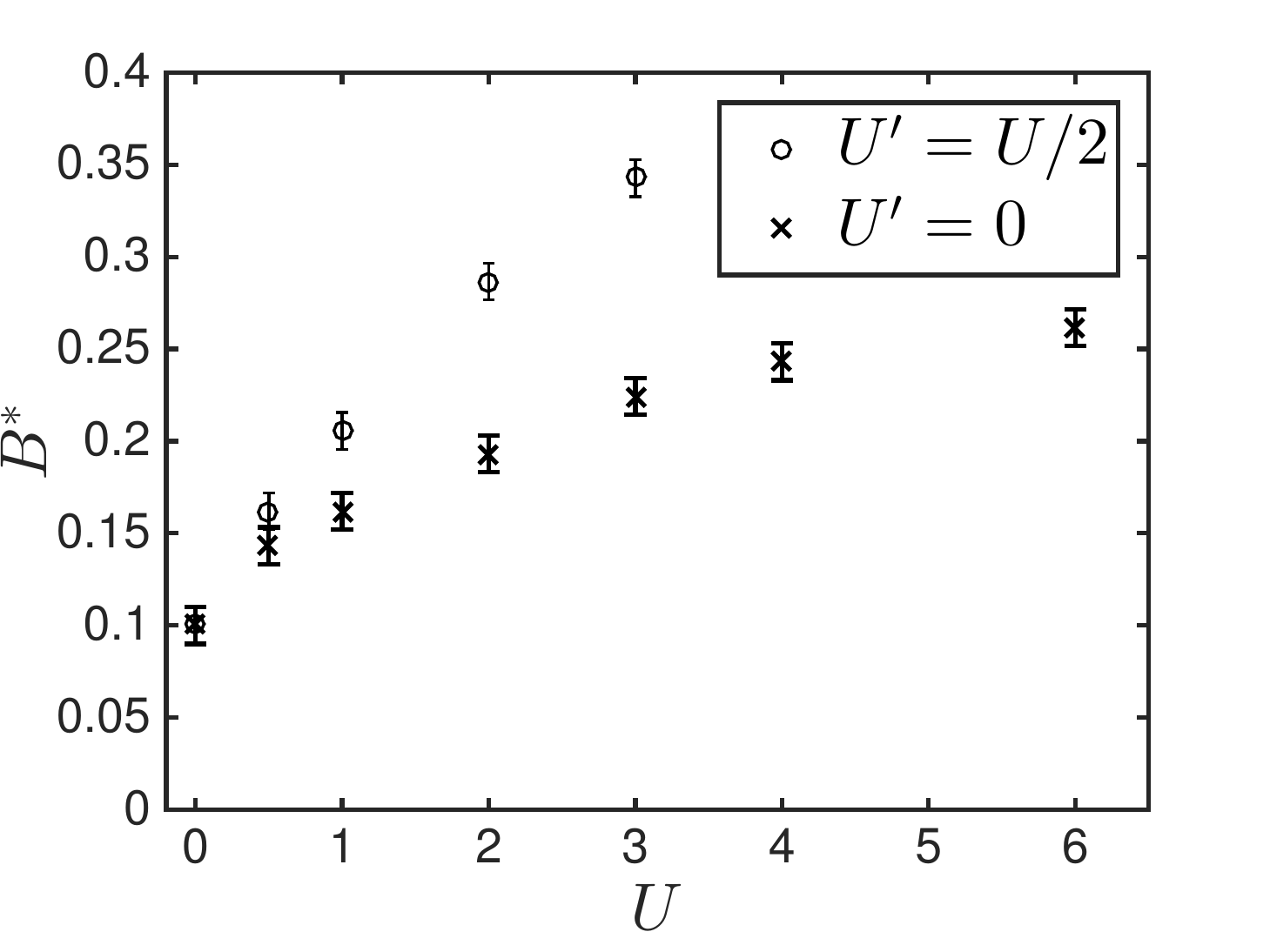}}}
  \caption{ (a) Spectral function $S_{cc^\dag}(k,\omega)$ for $n=0.5$,
    $\alpha=1$, $B=0.1$ and several values of $U$ (with $U'=U/2$,
    system size $L=128$ and $N=nL$ fermions).  The dotted red line is
    the Fermi energy, determined from the ground state energies as
    $E_F= (E(L,N+1)-E(L,N-1))/2$.  (b) SO gap parameter $B^*$ as
    obtained from the gap of size $2B^*$ in the spectral functions
    (shown until the MI phase sets in at $U=4$).  Results for $U'=0$
    are reported in \ref{us0}. }
  \label{fig:spec}
\end{figure}

We now move on to analyze the effect of e-e interaction on spectral
functions of the quantum wire. Results are shown in
figure~\ref{fig:spec}.  The data was obtained from evolving
Eq.~\eqref{eq:green} up to $t_{\text{max}}=25$ on a system with
$L=128$ sites. Using linear prediction, the time series was
extrapolated to $t_\mathrm{LP}=275$.  We note that at larger
interaction strengths the truncation error $\epsilon_\text{tot}(t)$
grows more rapidly with $t$ ~\footnote{For $m=1200$ states the total
  cumulative truncated weight $\epsilon_\text{tot}\leq 0.1$ for
  $U\leq 4$ (with $U'=U/2$), while $\epsilon_\text{tot}=0.25$ for
  $U=6$ (and $U'=3$).}. We have checked all our results for
convergence in $m$.  The spectrum in figure~\ref{fig:spec}~(a) at
$U=0.5$ exemplifies the spectral resolution of our approach.
Figure \ref{fig:spec} (a) shows the evolution of the spectral function
with the Coulomb interaction $U$, for $\alpha=1$, $B=0.1$, and at
$U'=U/2$ for which the two-particle backscattering is suppressed at
quarter filling. For comparison spectral function for $U'=0$ are shown
in \ref{us0}. With increasing interaction, the two noninteracting
branches get significantly broadened, but remain visible up to large
values of $U=6$.  Close to the Fermi energy, the spectral functions
remain sharp, as predicted by the LL theory.  The bandwidth increases
with the interaction, and so does the SO gap. This is shown in figure
\ref{fig:spec} (b).  Within the spiral LL approximation, the
correlation-induced enhancement of the gap in the is found to be
\cite{braunecker_nuclear_2009, braunecker_nuclear_2009_1,
  zamolodchikov_mass_1995, maier_nanowires_2014}
\begin{equation}
  2B^* =2B \xi^{1-\kappa/2}\;,
  \label{eq:bs}
\end{equation}
where $\kappa=K_\rho + K_\sigma^{-1}$ is determined from charge and
spin LL parameters $K_\rho$ and $K_\sigma$ and a correlation length
$\xi$.  For the noninteracting system, $K_\rho = K_\sigma = 1$ and
$B^* = B$. Repulsive interactions, for which 
$K_\rho < 1$ and $K_\sigma > 1$, lead to $\kappa < 2$. As a
consequence, the SO gap is enhanced by the interactions, which is
confirmed by figure \ref{fig:spec}~(b).

For $U=6$, the opening of a gap at $\pm k_\mathrm{F}$ indicates the
emergence of a Mott insulating (MI) phase at quarter filling
\cite{giamarchi2004quantum,mila_phase_1993,Essler_Weakly_2002,PhysRevLett.90.126401},
as described in section \ref{phase}.  We note that the MI phase is
absent for $U'=0$, see \ref{us0}.

\begin{figure}[t]
  \centerline{
    \subfloat[]{\includegraphics[width=0.408\textwidth]{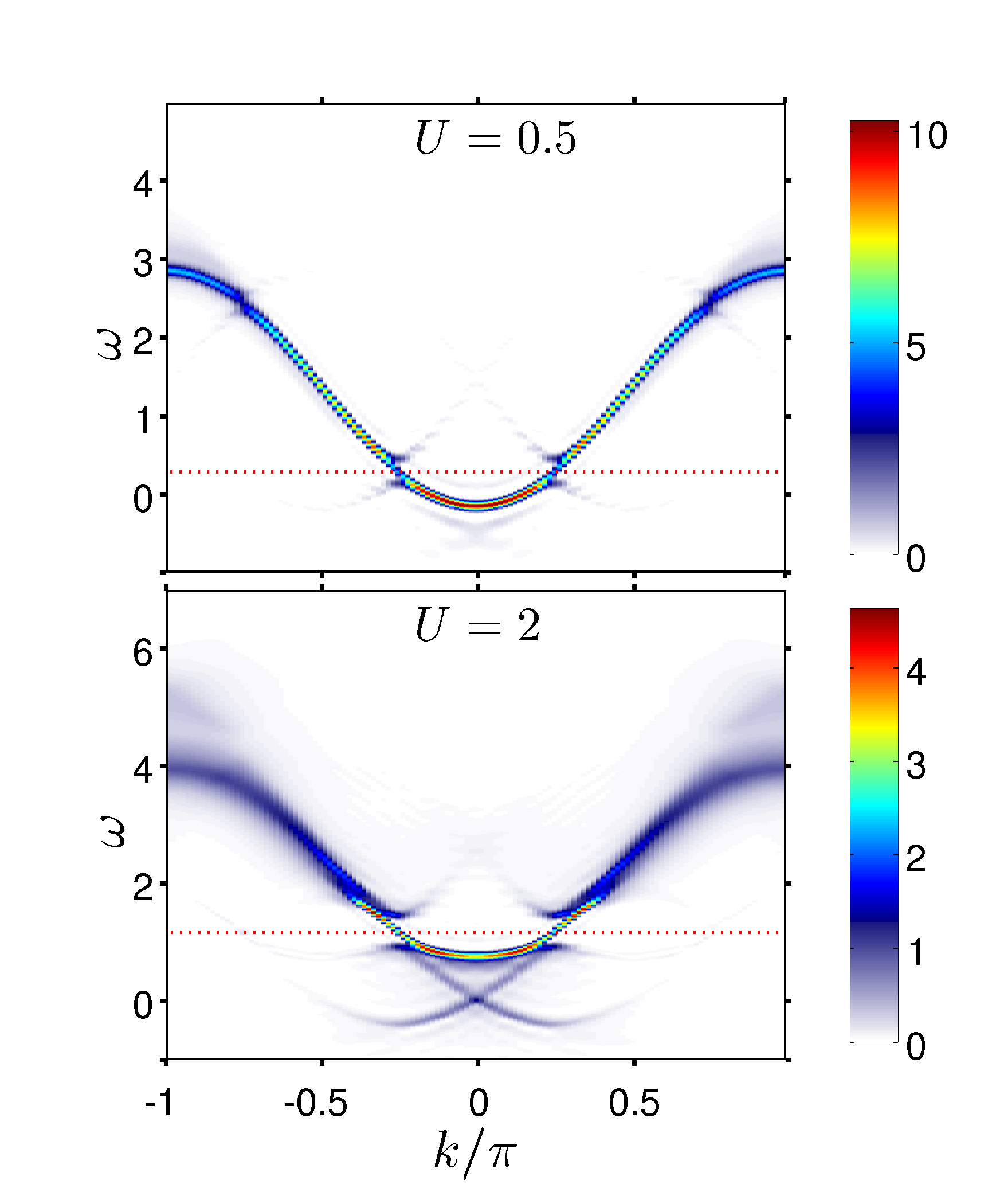}}
    \subfloat[]{\includegraphics[width=0.66\textwidth]{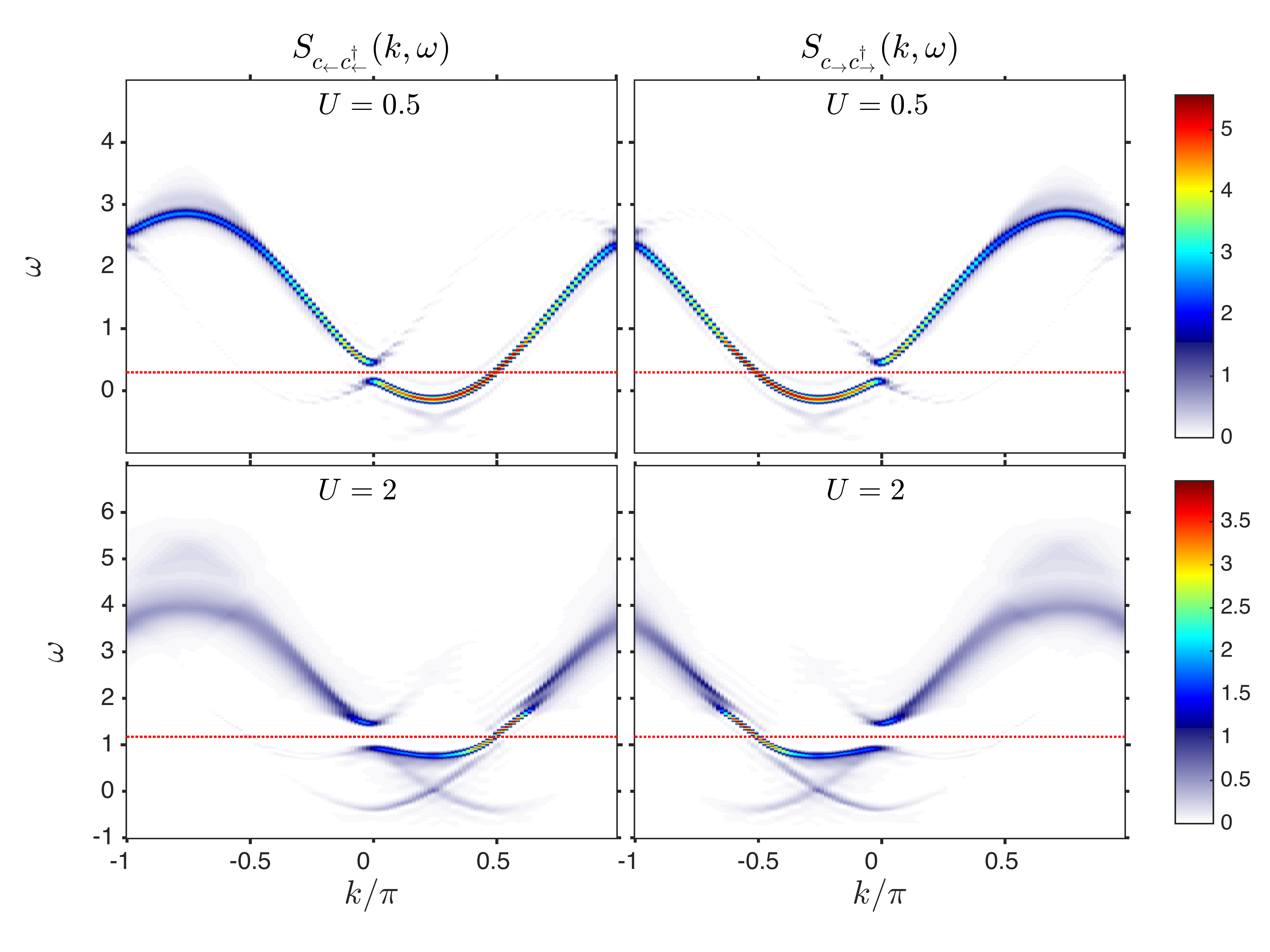}}}
  \caption{ (a) Spectral function in the gauge-transformed basis, at
    $n=0.5$, $\alpha=1$, $B=0.1$, $L=128$ and different values of $U$.
    (b) Spectral functions for individual spin components, parallel or
    antiparallel to the SO interactions, for the same couplings as in
    (a).  The spectra in (a) were obtained from the sum of the
    spectral functions of (b) shifted by $\pm k_\text{SO}$.  }
  \label{fig:specs}
\end{figure}

If the momentum shift of the SO interaction is removed by the gauge
transformation Eq.~\eqref{eq:gauge}~\footnote{This can be achieved by
  shifting $S_{c_\leftarrow c_\leftarrow^\dag}$ and
  $S_{c_\rightarrow c_\rightarrow^\dag}$ in momentum by
  $-k_\mathrm{SO}$, respectively $+k_\mathrm{SO}$, and summing them
  up.}, the similarity to the dispersion of the Hubbard model
facilitates the interpretation of the other features of the spectral
functions. Figure \ref{fig:specs} (a) shows that below the Fermi level
several dispersing branches appear, originating from the spin-charge
separation.  The distinctive main branch and the two weaker ones below
can be attributed to the collective spinon and holon excitations
\cite{PhysRevLett.92.256401,Schmidt_Fate_2010}, respectively. The
spectral weight of both increases with increasing interaction.

In figure \ref{fig:specs} (b) we show the spin-resolved components of
the spectral functions for the parallel and antiparallel direction
with respect to the SO interaction (see arrows in figure
\ref{fig:disp}), as determined by Eq.\eqref{eq:skw} for
$c^\dag_\ra = (c^\dag_\uparrow + i c^\dag_\downarrow)/\sqrt 2$ and
$c^\dag_\la = (c^\dag_\uparrow - i c^\dag_\downarrow)/\sqrt 2$.  These
directions are of particular interest, since they determine the
spin-dependent transport properties.
Each spin direction contains only a single Fermi point with
non-negligible spectral weight. Thus we find that the helical spin
order is robust with respect to electronic correlations and that spin
transport is still polarized.  Since $n=0.5$, the SO gap opens at
$k_\text F$ like for the spiral LL dispersion displayed in figure
\ref{fig:disp} (c).

Summarizing, the correlation effects observed for the microscopic
model are in agreement with the LL predictions of an enhanced SO gap
in the spectral functions and of the preservation of the helical spin
order with spin-dependent transport within the metallic 2F
phase~\cite{braunecker_nuclear_2009,braunecker_nuclear_2009_1,schuricht_spectral_2011,stoudenmire_interaction_2011}.

\subsection{Phase diagrams} \label{phase}
\subsubsection{Spiral and helical phases.} \label{2F_4F} In many
applications -- like Majorana wires, spin filters and Cooper pair
splitters -- it is crucial that the system is in the helical 2F
phase~\cite{oreg_helical_2010,lutchyn_majorana_2010,birkholz2008spin,streda_antisymmetric_2003,sato_cooper_2010}.
Therefore, and in order to guide experimental realizations, we
investigate the phase boundaries between the ordinary 4F phases and
the helical 2F phase.  In principle, the number of Fermi points can be
obtained by simply counting them in the spectrum.  However, the
calculation of spectral densities at many sets of parameters is
computationally rather expensive as compared to e.g. ground state
calculations.  Therefore, a reliable method to find the number of
Fermi points based only on ground state properties is favorable.
\begin{figure}[t]
  \centerline{
    \subfloat[]{\includegraphics[clip=true,trim=0.2cm 0cm 0.6cm 0cm,width=0.5\textwidth]{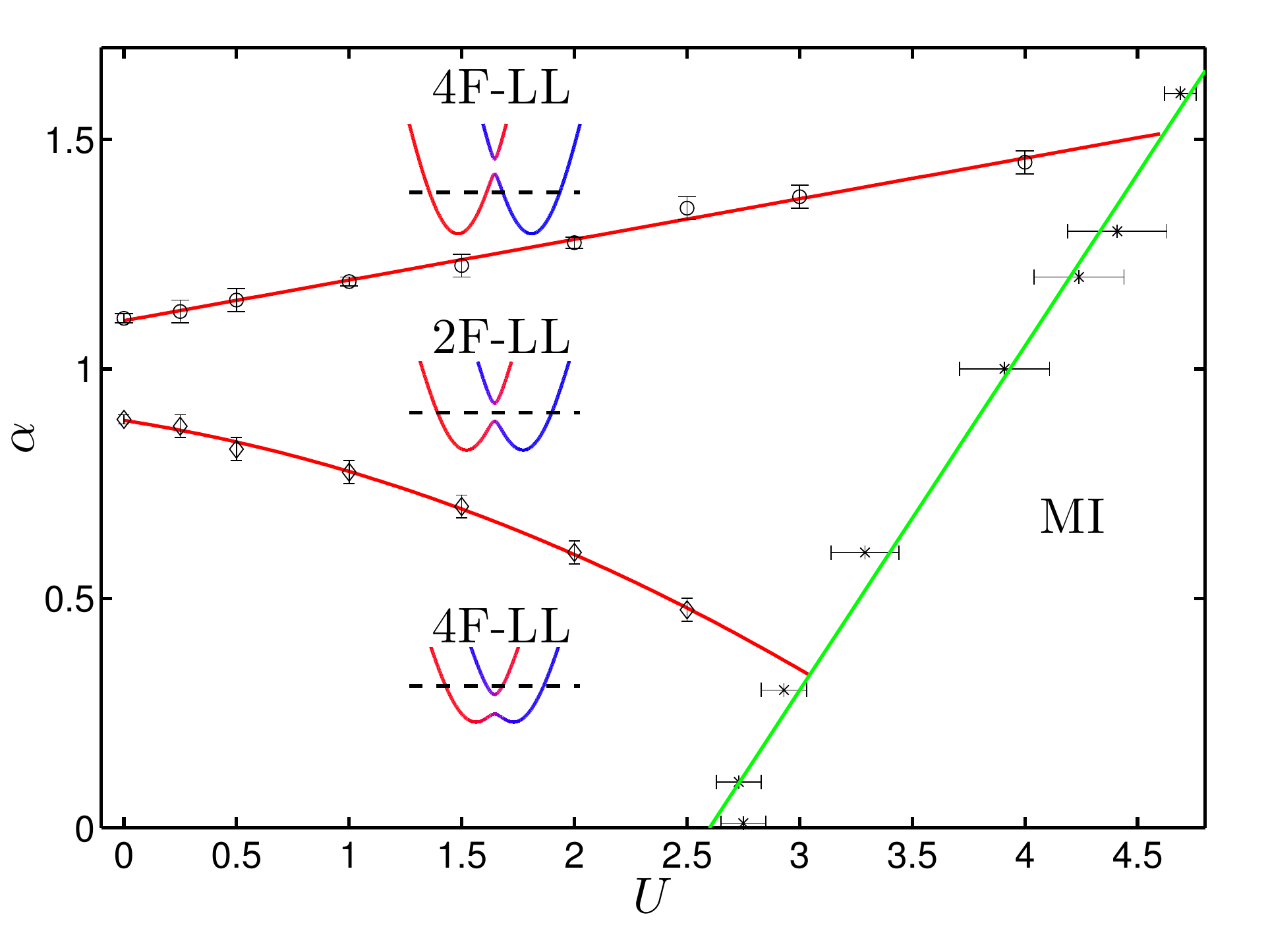}}\\
    \raisebox{0.02cm}[0pt]{\subfloat[]{\includegraphics[clip=true,trim=0.2cm
        0cm 0.6cm 0cm,width=0.5\textwidth]{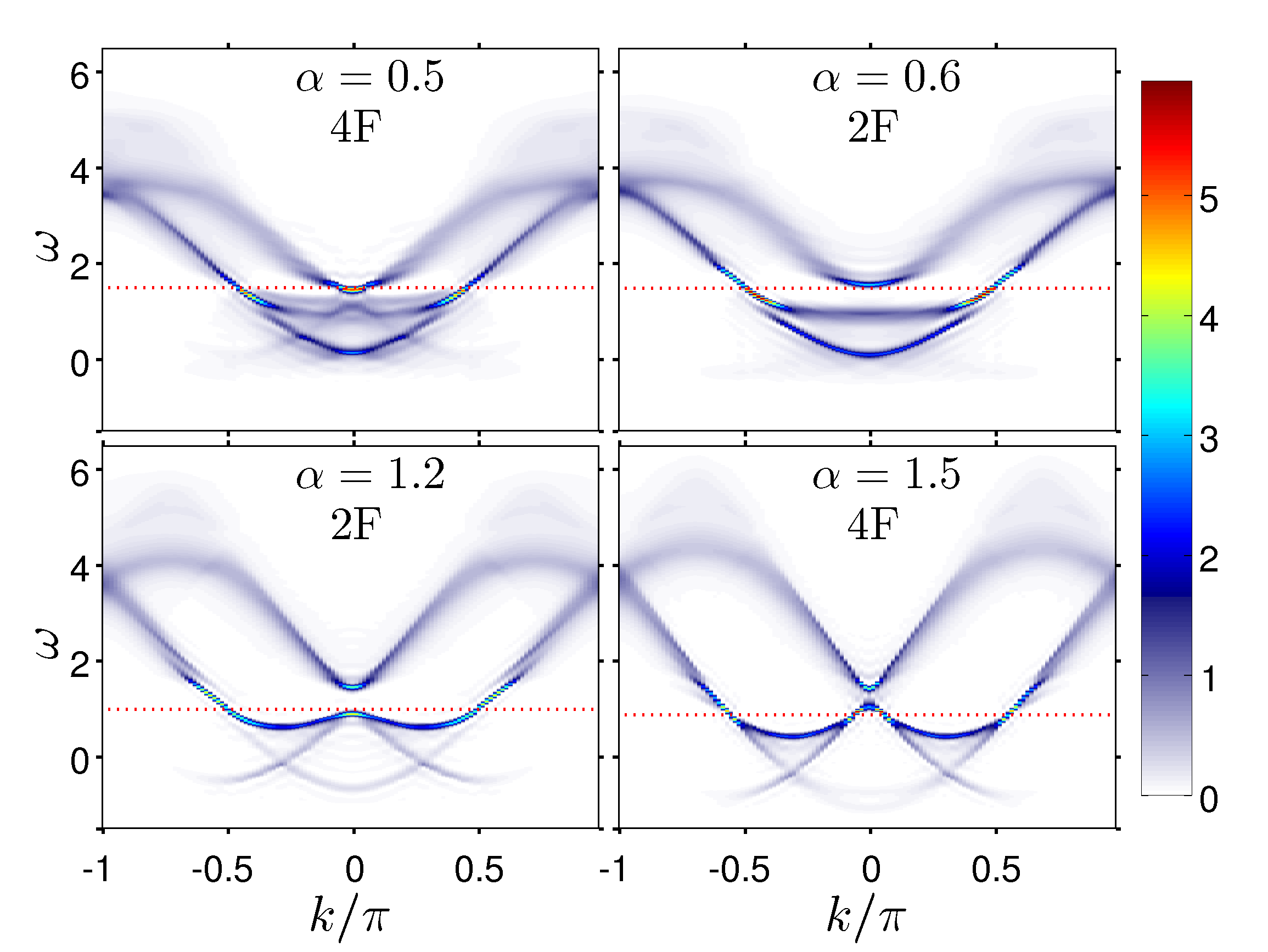}}}}
  \caption{(a) Phase diagram ($\alpha$,$U$), for $n=0.5$, $B=0.1$ and
    $L=128$.  The phase boundaries between the 2F and 4F phases are
    obtained from the discontinuity in $K_\rho$ (Eq.\
    \eqref{eq:krho}). The phase boundary of the MI was obtained from
    the closing of the two particle charge gap in DMRG.  The red and
    green lines are polynomial fits (using second order for the lower
    red line and first order for the upper red and green lines).  (b)
    Spectra in the 4F and 2F phases at $U=2$ for different SO
    interactions $\alpha$.  }
  \label{fig:phase_alpha}
\end{figure}
To this end, we use a method based on the calculation of the static
density-density correlation function
$C_{\hat n\hat n}(r)\equiv \bra{0}\hat n_r\hat n_0\ket{0}$.  Within
the 2F phase, our system is described by a spinless fermion LL, and
the asymptotic behavior ($1 \ll r \ll L$) of the density-density
correlations is given by \cite{ejima_tomonaga-luttinger_2005}
\begin{equation}
  C_{\hat n \hat n}(r) \sim - \frac{K_{\rho}}{2(\pi r)^2}+A\frac{\cos(2k_\text F r)}{r^{1+K_\rho}}\ln^{-\frac{3}{2}}(r)+... ,
  \label{eq:cnn}
\end{equation}
with a model-dependent constant $A$. The second expression contributes
only logarithmically and is neglected. After a Fourier transformation,
$K_\rho$ can be obtained from the derivative at $k=0$ in the
thermodynamic limit, or from a finite size extrapolation of
$L\rightarrow \infty$:
\begin{equation}
  K_\rho =\lim_{L \ra \infty} L \, \tilde C_{\hat n \hat n}(k=\frac{2\pi}{L}).
  \label{eq:krho}
\end{equation}

In the 4F phase on the other hand, the low energy physics is no longer
described by the simple LL for spinless fermions. Nevertheless, the
leading large-distance behavior of the density-density correlations is
still quadratic, namely $K_\rho/(\pi r)^2$. Thus the change in the
prefactor of $1/r^2$, by a factor of $2$, can be used to distinguish
the 2F and 4F phases.

\begin{figure}[t]
  \centerline{ \subfloat[]{\includegraphics[clip=true,trim=0.2cm 0cm
      0.6cm 0cm,width=0.5\textwidth]{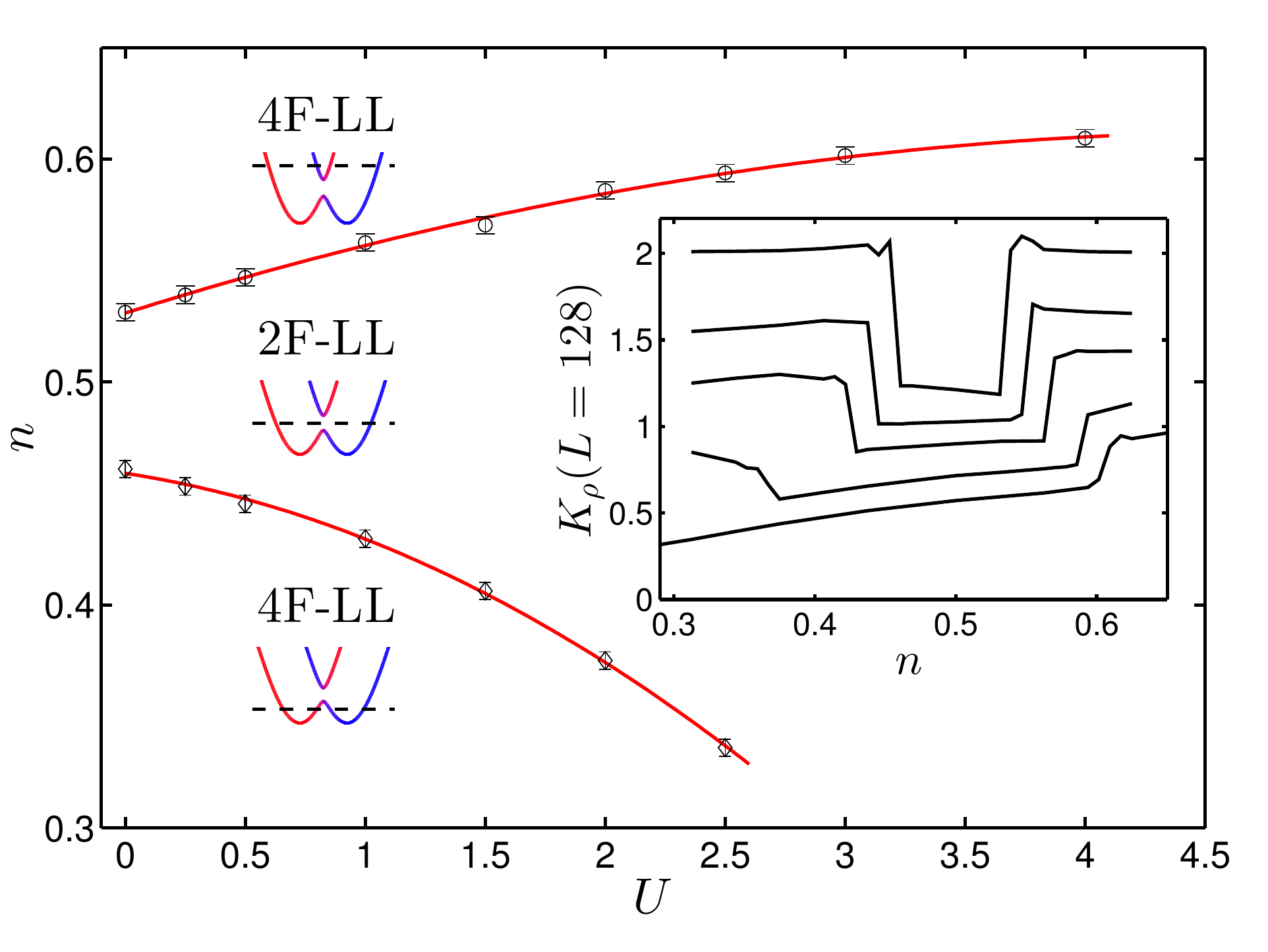}}
    \raisebox{0.02cm}[0pt]{\subfloat[]{\includegraphics[clip=true,trim=0.2cm
        0cm 0.6cm 0cm,width=0.5\textwidth]{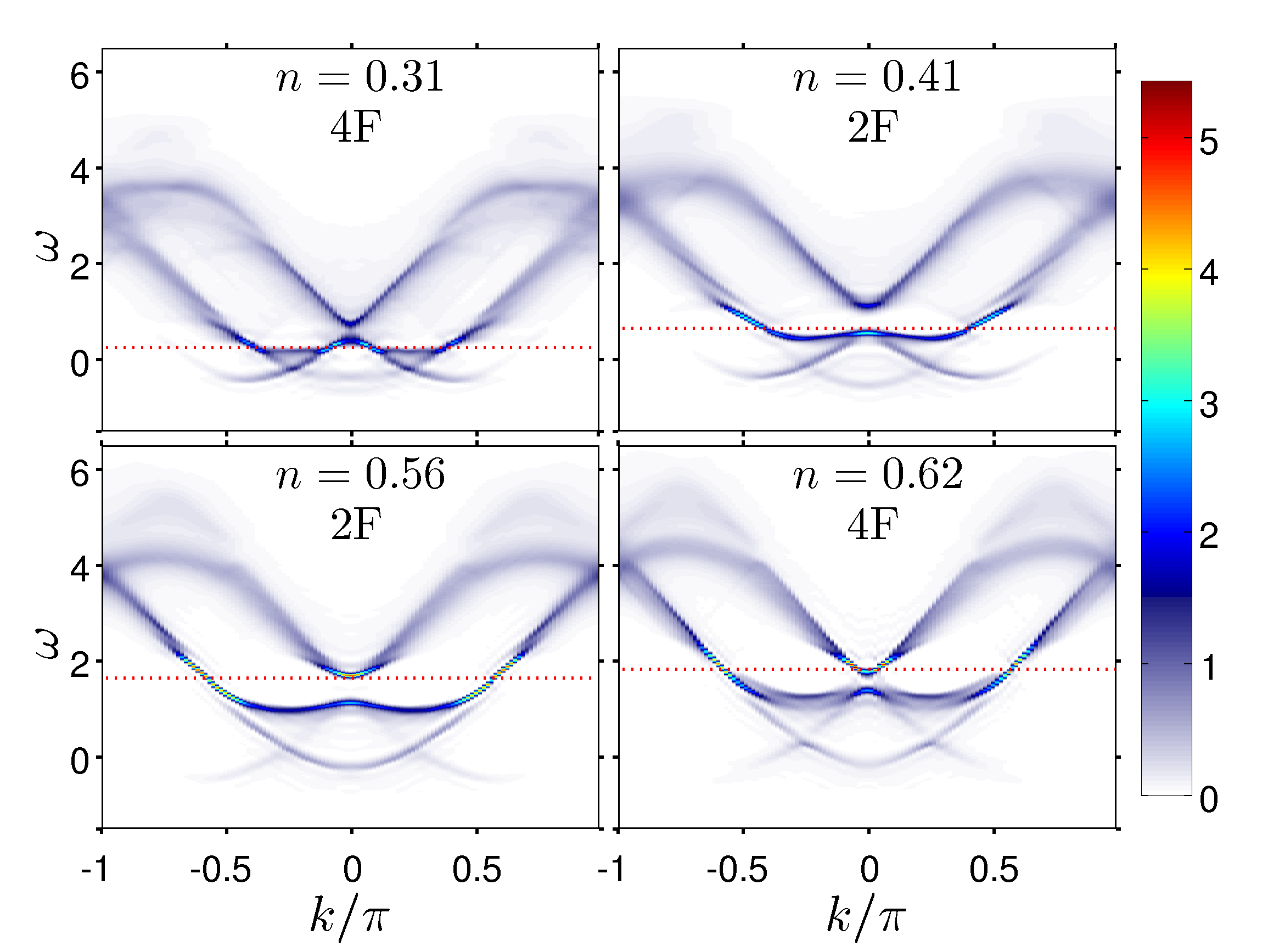}}}}
  \caption{(a) Phase diagram ($n$,$U$), for $\alpha=1$, $B=0.1$ and
    $L=128$.  The phase boundaries between the 2F and 4F phases are
    obtained from the discontinuity in $K_\rho$ (see Eq.\
    \eqref{eq:krho}).  The inset shows the behavior of $K_\rho$
    for different values of $U=0$, $0.5$, $1$, $2$, $3$ from top to
    bottom.  The red lines are polynomial fits of second order.  (b)
    Spectra in the 4F and 2F phases at $U=2$, for different fillings
    $n$.  }
  \label{fig:phase_mu}
\end{figure}

The phase diagrams obtained in this way are shown in figures
\ref{fig:phase_alpha} (a) and \ref{fig:phase_mu} (a). For the phase
boundaries between 2F and 4F phases (red lines in figures
\ref{fig:phase_alpha} (a) and \ref{fig:phase_mu} (a)) we used a system
size of $L=128$ (the results for $L=256$ are indistinguishable from
$L=128$). In figure \ref{fig:phase_alpha} (a), the phase boundary to
the MI (green line) was obtained from the closing of the two particle
charge gap (see below). The inset in figure \ref{fig:phase_mu} (a)
shows the behavior of the LL parameter $K_\rho$ as determined by
Eq. \eqref{eq:krho}, with the jump at the phase boundaries between the
2F and 4F phases.  Except for $U=0$, the actual jump of $K_{\rho}$
between the two phases is in general smaller than a factor of
2~\footnote{This is also true after a finite size extrapolation, as it
  was performed e.g. in Fig.~\ref{fig:mott}~(a)}.

The phase boundaries in $(\alpha,U)$ (figure \ref{fig:phase_alpha})
were obtained at quarter filling $n=0.5$, and in $(n,U)$ at $\alpha=1$
(figure \ref{fig:phase_mu}).  In the noninteracting case, the
boundaries of the 2F phase in $(\alpha, U=0)$ (at $n=0.5$) are at
\begin{equation}
  \sqrt{t^2-2Bt}<\alpha < \sqrt{t^2+2Bt}\;,
  \label{eq:abound}
\end{equation}
and in $(n,U=0)$ for fixed $\alpha$ at
\begin{equation}
  \arccos\left(\frac{t^2-\alpha^2+2Bt}{t^2+\alpha^2}\right)<\pi n<\arccos\left(\frac{t^2-\alpha^2-2Bt}{t^2+\alpha^2}\right)\;.
  \label{eq:nbound}
\end{equation}
The phase boundaries at $U=0$ obtained from Eqs.~\eqref{eq:abound}
and \eqref{eq:nbound} lie within the error bars of the numerically
obtained ones in figure \ref{fig:phase_alpha} and \ref{fig:phase_mu}.

Remarkably, we find that with interactions the parameter range of the
2F phase gets greatly enhanced \cite{stoudenmire_interaction_2011},
which is in accordance with the interaction-enhanced gap we already
found for the spectral functions in figure \ref{fig:spec} (a).
Examples for spectral functions in the different phases are given in
figures \ref{fig:phase_alpha} (b) and \ref{fig:phase_mu} (b). We show
the spectral functions in the proximity of the phase boundaries.  Note
how the SO gap gets enhanced as soon as the Fermi energy slips inside
the 2F phase.  Therefore, we predict repulsive electron-electron
interactions to be beneficial to applications depending on the 2F
phase, like Majorana zero
modes~\cite{stoudenmire_interaction_2011,thomale_chebyshev_2013,sela_majorana_2011,hassler_majorana_2012}
or spin-filters. 

For both phase diagrams (figures~\ref{fig:phase_alpha} and
\ref{fig:phase_mu}), the jump in $K_\rho$ (only shown in
figure~\ref{fig:phase_mu}) is no longer visible in the strong coupling
regime, indicating that the LL picture breaks down there.


\subsubsection{Mott phase.}
\begin{figure}[t]
  \center
  \subfloat[]{\includegraphics[width=0.5\textwidth]{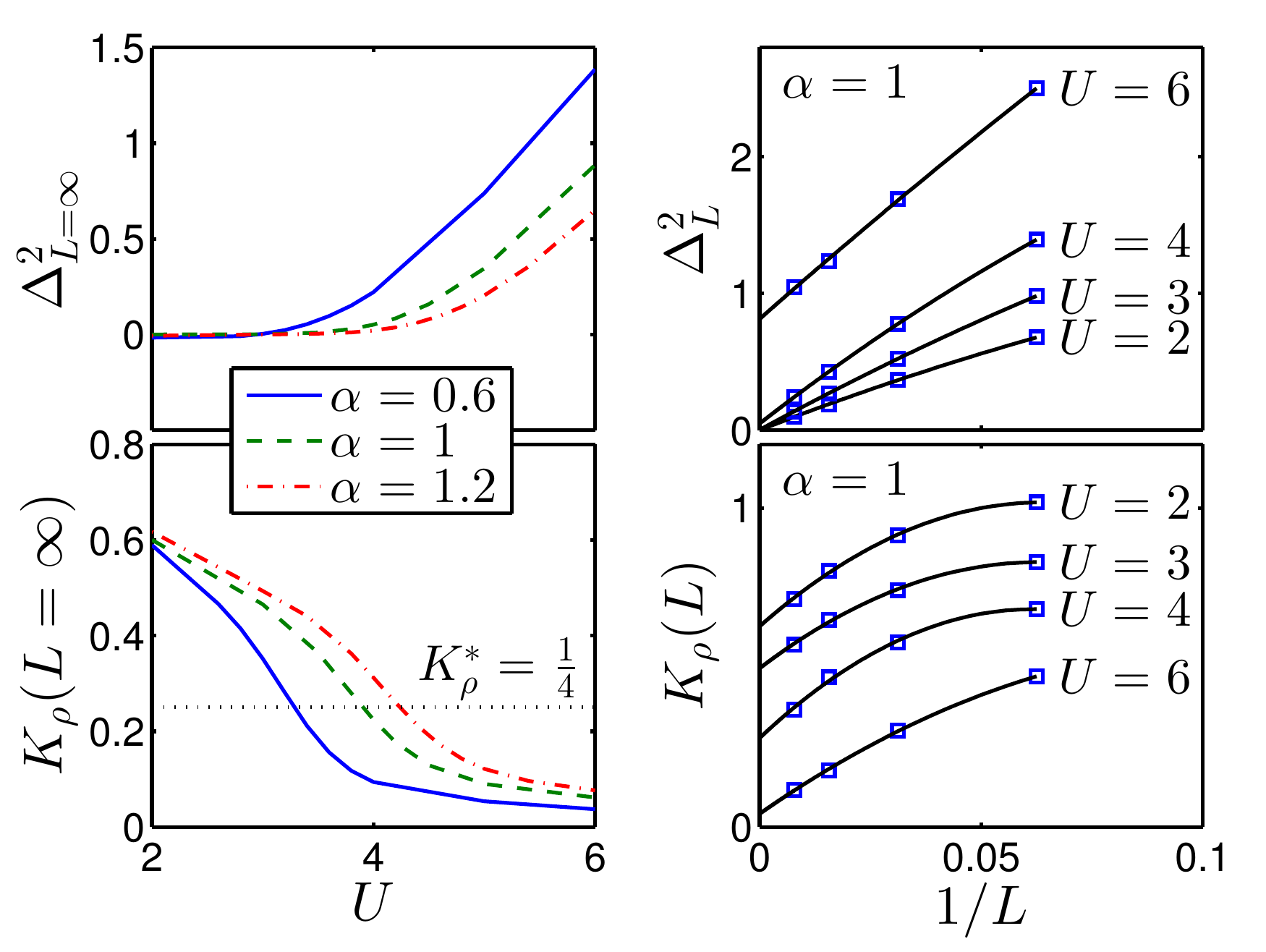}}
  \subfloat[]{\includegraphics[width=0.5\textwidth]{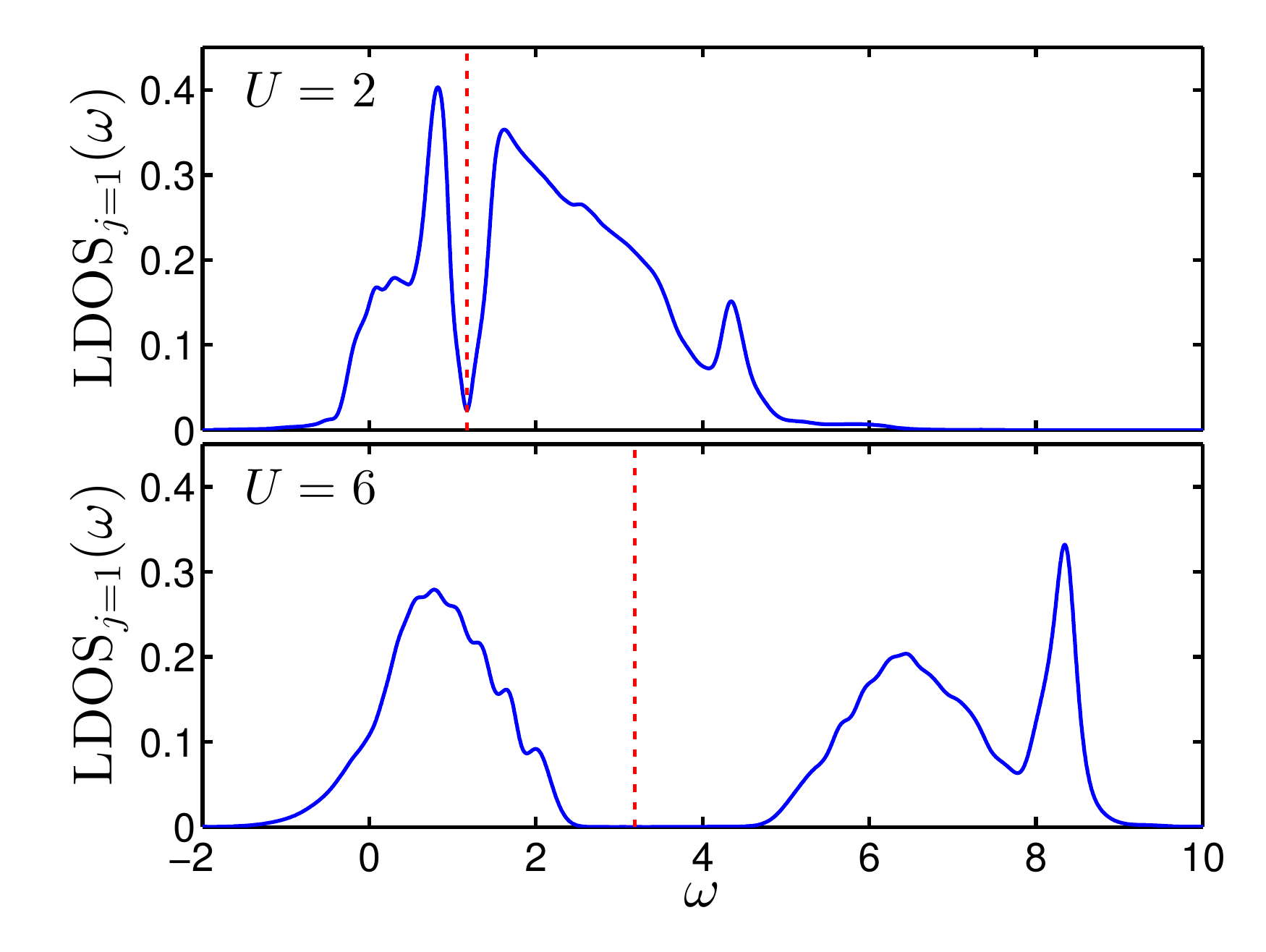}}
  \caption{(a) The boundary to the MI as a function of $U$ was
    determined from the charge gap (upper left panel)
    $\Delta^2_{L=\infty}$ as well as from the critical value of
    $K_\rho^*=0.25$ (lower left panel).  The finite-size
    extrapolations are shown for $\alpha=1$ in the right panel (using
    system sizes $L=\{16, 32, 64, 128$\}).  (b) Local density of
    states at the chain boundary for $U=2$ (in the LL phase) and $U=6$
    (in the MI phase), both at $\alpha=1$. The dotted red line is the
    Fermi energy.  For all panels the other parameters are $n=0.5$,
    $B=0.1$ and $L=128$.  }
  \label{fig:mott}
\end{figure}

At quarter filling, the strong coupling phase is a Mott Insulator
\cite{giamarchi2004quantum, mila_phase_1993}. To detect the MI phase
transition in our model, we use two different approaches, to be
detailed in the following. The first one is via the calculation of the
two-particle (charge) excitation gap above the ground state,
\begin{equation}
  \Delta_L^2=\frac{1}{2}\bigl[E(L,N+2)+E(L,N-2)-2E(L,N)\bigr]\;,
  \label{eq:tpe}
\end{equation}
with $N$ being the absolute filling $N=n L$. In the absence of pairing
effects, the two particle and the single particle excitation gap will
scale to the same value in the thermodynamic limit, but the two
particle excitation gap is robust to even/odd effects. $\Delta_L^2$ is
calculated from ground state energy of three different $DMRG$ runs,
one for each total filling $N+2,N$ and $N-2$.  Complementary to this,
we use the evolution of $K_{\rho}$ with increasing $U$ to detect the
MI phase transition. For quarter filling, there is a critical value
$K_\rho^*=0.25$ below which the system is in a MI state
\cite{giamarchi2004quantum}.  Note that the value of $K_\rho$ from
Eq.~\eqref{eq:krho} in the 4F phase, Eq.~\eqref{eq:krho}, differs by a
factor of two as compared to the 2F case, see the discussion in
Sec.~\ref{2F_4F}.  For both approaches, the values in the
thermodynamic limit $L\rightarrow\infty$ were obtained by a polynomial
fit in $1/L$, as shown in the right column of figure \ref{fig:mott}
(a).  The left column in figure \ref{fig:mott} (a) shows the results
in the thermodynamic limit $L=\infty$.  The phase boundaries obtained
from the two approaches agree within the error bars and are shown as
the green line in figure \ref{fig:phase_alpha} (a).  Note how the
phase boundary to the MI moves to larger values of $U$ with increasing
SO interaction. This can be understood from the contribution of the SO
interaction to the kinetic energy of the system, which lowers the
effective interaction at fixed $U$ and drives the system away from the
MI.

The charge gap in the MI is also clearly visible in the local density
of states LDOS$_{j=1}(\omega)$ at the left chain boundary, as shown in
figure \ref{fig:mott} (b).  In the LL phase we find a pronounced
interaction-induced suppression of the local spectral weight
\cite{Schoenhammer_Boundary_2000, Meden_Luttinger_2000}, while in the
Mott phase a gap opens around the Fermi energy. The high-energy peak
in the LDOS corresponds to the upper Hubbard band, see also figure
\ref{fig:spec}.

Away from quarter filling, a strong coupling region emerges at strong
interactions which is further examined in \ref{cdw}.

\subsection{Breathers, structure factors and optical conductivity}
\label{breather}

\subsubsection{Breather bound states.}
In the mathematical formulation of the spiral LL, the gapped modes are
described by a sine-Gordon model (SGM). The elementary excitations in
the SGM are solitons and antisolitons with the mass $B^*$ of half the
SO gap.  In the attractive regime of the SGM (which amounts to
repulsive interactions in the spiral LL), additional
soliton-antisoliton bound states appear, which are so-called breather
states \cite{shifman2005fields, gogolin2004bosonization}.  In the
spiral LL theory their masses are given by
\cite{schuricht_spectral_2011}
\begin{equation}
  \Delta_l = 2 B^* \sin\left(\frac{l \pi \kappa}{8-2\kappa}\right)\;,
  \label{eq:breather}
\end{equation}
with $l=1,\dots,l_\text{max}$ and
$l_\text{max}=\text{int}( 4/\kappa -1)$ different breathers (for
$\kappa < 2$, with $\kappa=K_\rho + 1/K_\sigma$).

\begin{figure}[t]
  \center
  \includegraphics[width=0.5\textwidth]{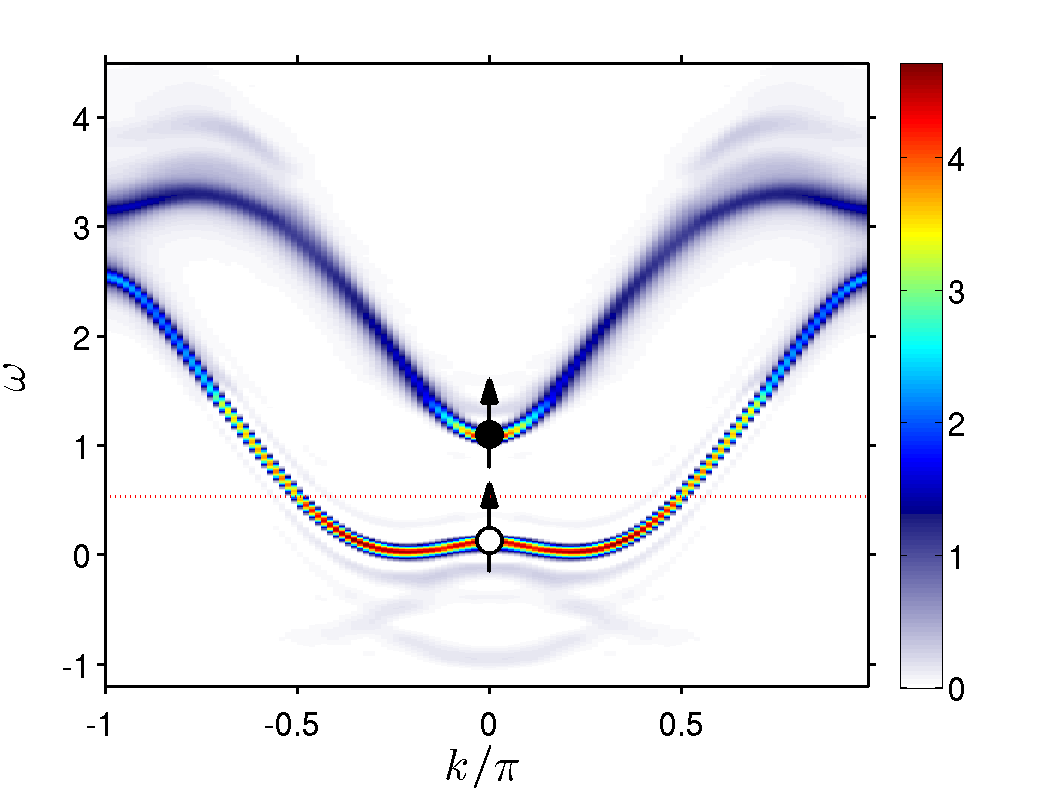}
  \caption{Illustration of breather bound states on the gapped modes,
    consisting of a particle excitation (soliton) on the upper branch
    and a hole excitation (antisoliton) on the lower branch. The
    spectral function shown here was obtained for $n=0.5$, $\alpha=1$,
    $U=1$, $B=0.3$ and $L=128$.  }
  \label{fig:breatherdisp}
\end{figure}

In our lattice model the breathers correspond to bound states of a
particle excitation (soliton) and a hole excitation (antisoliton) on
the gapped modes (see figure \ref{fig:breatherdisp} for an
illustration).  They are charge neutral but carry a positive
$\hat S^z$ magnetization. The electron and the hole are bound together
by the interaction, with an energy smaller than the SO gap $2B^*$.  In
the real-time evolution, breathers oscillate back and forth around
their ``center of mass'', which motivates their naming.  The breather
contributions to one-particle Green functions, like the spectral
function or the local density of states, are found to be negligible
\cite{schuricht_spectral_2011}.  However, the breathers strongly
couple to the current density $\hat J$, hence the optical conductivity
\begin{equation}
  \sigma(\omega>0)=\frac{\text{Im}(S_{\hat J \hat J}(k=0,w>0))}{L\omega}
  \label{eq:oc}
\end{equation}
is an excellent choice for observing breather bound states.

In the following, we will compare the optical conductivity
$\sigma(\omega)$ of our microscopic model to the optical conductivity
$\sigma_{SLL}(\omega)$ of the spiral LL, which has been calculated in
Ref.~\cite{schuricht_spectral_2011}. We will determine the necessary
parameters $B^*$, $K_\rho$, $K_\sigma$ and the charge and spin
velocities $v_\rho$ and $v_\sigma$ for the calculation of
$\sigma_{SLL}(\omega)$ from our microscopic model.  This comparison
will serve as an important test for the consistency and validity of
the spiral LL approach and our numerical calculations.

Since we are now focusing on the physical behavior taking place at
energies smaller than the SO gap, it is advantageous to use a stronger
Zeeman field of $B=0.3$ for a wider gap.  The other parameters are as
before, $\alpha=1$ and $n=0.5$, different interactions $U$, and
$U'=U/2$.  The higher magnetic field $B = 0.3$ causes the MI phase to
set in earlier (from $U \ge 2$ on there are hints of a quarter filled
MI order).  We extracted the values of the spiral LL parameters from
several different observables.

\subsubsection{Local density of states.}
\begin{figure}[t]
  \includegraphics[width=1\textwidth]{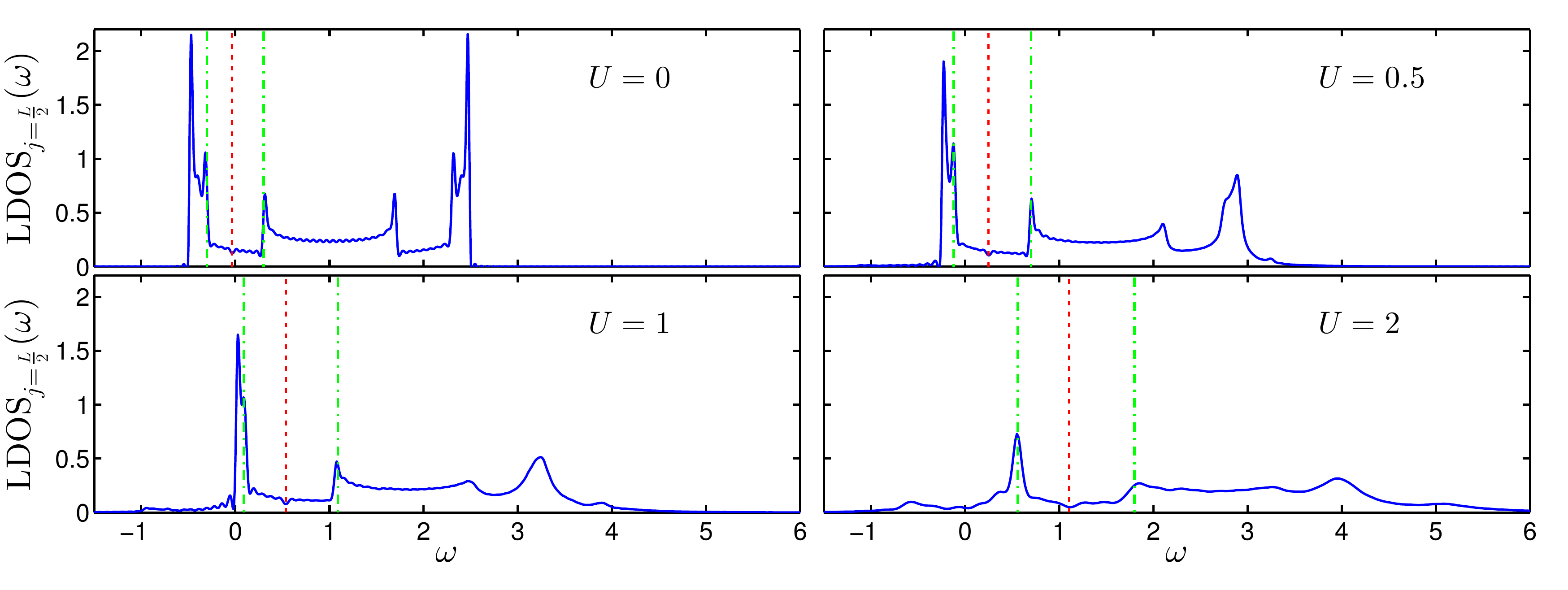}
  \caption{Local density of states in the middle of a $L=128$ site
    chain for $n=0.5$, $\alpha=1$, $B=0.3$ and different values of
    $U$. The dashed red line is the Fermi energy, whereas the
    dot-dashed green lines mark the boundaries of the SO gap.}
  \label{fig:ldos}
\end{figure}

The renormalized size $2B^*$ of the SO gap was extracted from the
local density of states (LDOS) in figure \ref{fig:ldos}. Due to the
local nature of the LDOS we were able to use longer time evolutions
than for the momentum-resolved spectra in the same systems, reaching
from $t_{\text{max}}=30$ up to $t_{\text{max}}=90$ depending on the
interaction.  The small scale oscillations visible in figure
\ref{fig:ldos} at lower values of the interaction are consistent with
the energy spacing for an $L=128$ site system.  Results for $B^*$ are
shown in table \ref{tab:par}.

\begin{table}[htb]
  \begin{center}
    \begin{tabular}{@{}llllllc}
      \br
      $U$ & $K_\rho$ & $K_\sigma$  & $v_\rho$ & $v_\sigma$ & $B^*$  & $\Delta_1$ \\
      \mr
      0   & 1.00     & 1.00  & 1.0     & 1.0       & 0.30   & --- \\
      0.5 & 0.87     & 1.20  & 1.0     & 1.0       & 0.41   & 0.75 (0.75) \\
      1   & 0.76     & 1.28  & 1.0     & 1.1       & 0.50   & 0.83 (0.83) \\
      2   & 0.59     & 1.29  & 1.2     & 1.3       & 0.62   & 0.89 (0.90) \\
      \br
    \end{tabular}
    \caption{\label{tab:par}Parameters $K_\rho$, $K_\sigma$, $v_\rho$,
      $v_\sigma$ and $B^*$ as used for the field-theoretical
      calculations, and the mass of the first breather
      $\Delta_1$. $K_\sigma$ was extracted from the energy of the
      first breather peak $\Delta_1$ using Eq.~\eqref{eq:breather},
      where $\Delta_1$ was obtained from the position of the peak in
      figure~\ref{fig:ocfb}. The value in parentheses is obtained from
      the oscillation frequency in figure~\ref{fig:breather}.  The
      estimated uncertainties in these values are about 5\% (10\% for
      the velocities).  }
  \end{center}
\end{table}

\subsubsection{Structure factors.}
\begin{figure}[t]
  \includegraphics[width=1\textwidth]{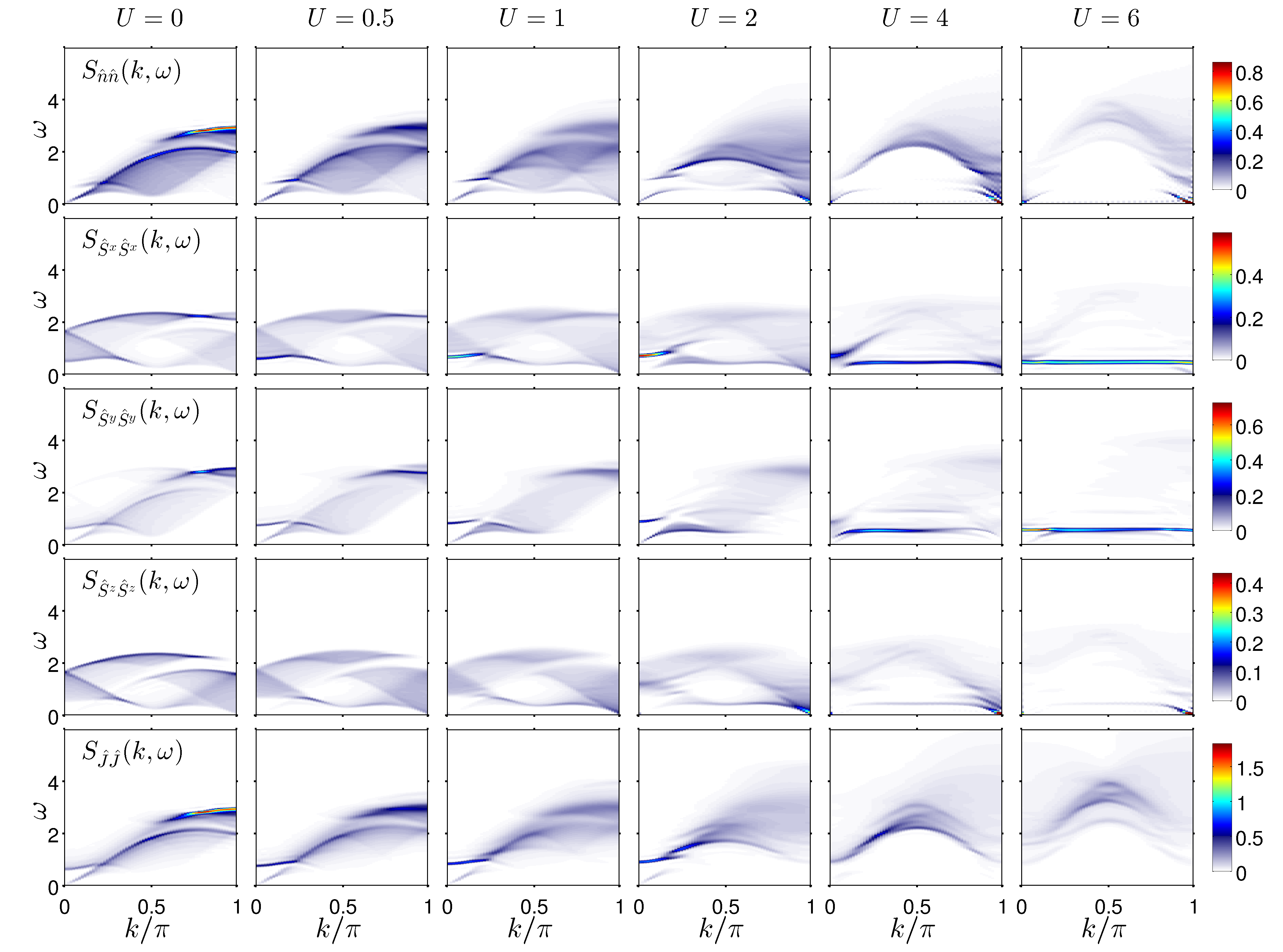}
  \caption{Charge, spin and current structure factors
    $ S_{\hat n \hat n}(k,\omega)$,
    $ S_{\hat S^x \hat S^x}(k,\omega)$,
    $ S_{\hat S^y \hat S^y}(k,\omega)$,
    $ S_{\hat S^z \hat S^z}(k,\omega)$ and
    $ S_{\hat J \hat J}(k,\omega)$, for $n=0.5$, $\alpha=1$, $B=0.3$,
    $L=128$ and different values of $U$.
    $ S_{\hat n \hat n}(k=\pi,\omega=0)$ and
    $ S_{\hat S^z \hat S^z}(k=\pi,\omega=0)$ diverge in the MI phase
    ($U > 2$) and exceed the color range, with maximal values in our
    analysis of $1.0$, $20.4$, $42.4$ for
    $ S_{\hat n \hat n}(k,\omega)$ and of $0.5$, $4.8$, $7.9$ for
    $ S_{\hat S^z \hat S^z}(k,\omega)$ at $U=2$, $4$, $6$
    respectively.  }
  \label{fig:structfac}
\end{figure}

We obtain the velocities $v_\rho$ and $v_\sigma$ from the
corresponding structure factors $ S_{\hat n \hat n}(k,\omega)$ and
$ S_{\hat S^y \hat S^y}(k,\omega)$.  The structure factors are shown
in figure~\ref{fig:structfac}.  Since the spectra are symmetric,
values for $k<0$ are not shown. Near $k=\omega=0$, the dispersions are
approximately linear (at least for the considered $U \le 2$) and the
velocities were obtained from fits to their
slopes. Table~\ref{tab:par} shows that for the interactions
considered, $v_\rho \approx v_\sigma$.

The structure factors themselves contain very interesting physics.  In
the weak coupling regime, the nearly linear dispersions starting at
$k=\omega=0$ visible in $ S_{\hat n \hat n}(k,\omega)$,
$ S_{\hat S^y \hat S^y}(k,\omega)$ and $ S_{\hat J \hat J}(k,\omega)$
result from the ungapped modes of the Hamiltonian.  The approximately
quadratic dispersion at $\omega = \Delta_1$ and $k=0$ in
$ S_{\hat S^x \hat S^x}(k,\omega)$, $ S_{\hat S^y \hat S^y}(k,\omega)$
and $ S_{\hat J \hat J}(k,\omega)$ corresponds to the breather
modes. Note that their dispersion starts at an energy $\Delta_1$,
which is smaller than the SO $2B^*$ gap obtained from the
single-particle spectra.  The breather dispersion gains more spectral
weight with increasing interaction but smears out in the MI phase from
$U \ge 2$ on.  Unlike spin charge separation, which also gives rise to
two low energy modes \cite{Pereira_Spin-Charge_2010}, the modes here
are not separated because of interactions but due to the interplay of
Zeeman field and SO interaction (each mode is a combination of spin
and charge degrees of freedom).

In contrast to an ordinary MI, at quarter filling there is no
antiferromagnetic order at $k=\pi/2$ in the spin structure factors.
Instead, the Zeeman field induces a finite magnetization. Therefore,
$ S_{\hat S^z \hat S^z}(k,\omega)$ and $ S_{\hat n \hat n}(k,\omega)$
show the same diverging behavior at $k=\pi$ and $\omega = 0$ for
strong interactions.  Deep in the MI phase,
$S_{\hat S^x \hat S^x}(k,\omega)$ and
$S_{\hat S^y \hat S^y}(k,\omega)$ consist almost solely of a constant
energy level at $\omega = 2B$, indicating that all movement of the
spins freezes except for a simple precession around the Zeeman field.
$S_{\hat n \hat n}(k,\omega)$ and $S_{\hat S^z \hat S^z}(k,\omega)$
are dominated by the Mott instability for strong interactions.  As
soon as the interactions are turned on, spectral weight at $\omega =0$
and $k=2k_\mathrm{F}=\pi$ accumulates, indicating the charge order of
the quarter filled MI.  In the MI phase, $S_{\hat n \hat n}(k,\omega)$
and also $S_{\hat S^z \hat S^z}(k,\omega)$ diverge at this point.

The LL parameter $K_\rho$ at $B=0.3$ was obtained from the static
density-density correlations, as described in section \ref{phase}.
We used system sizes from $L=16$ up to $L=256$ and extrapolated to
$L=\infty$ by a fourth order polynomial fit in $1/L$. The spin
parameter, $K_\sigma$, is known to be very susceptible to finite size
corrections \cite{ejima_dmrg_2009}. When we applied the same method to
$K_\sigma$, we found the extrapolation to the thermodynamic limit to
be unreliable, since it turned out to be nonmonotonic in $1/L$. Since
all other parameters for the field theory are already fixed, we choose
to obtain $K_\sigma$ with Eq.~\eqref{eq:breather} from the energy
of the breather peak in the optical conductivity $\sigma(\omega)$,
which we discuss now.

\subsubsection{Optical conductivity.}
According to Eq.~\eqref{eq:oc} the optical conductivity can be
obtained from $S_{\hat J \hat J}$. For our model, the usual Hubbard
current operator has to be adapted in order to include the SO interaction. It
then reads as follows for the current between the sites $j$ and $j+1$
\begin{equation}
  \begin{aligned}
    \hat J_{j} = i\left[-\frac{t}{2}\sum_\sigma (c_{j\sigma}^\dag
      c_{j+1\sigma}-c_{j+1\sigma}^\dag c_{j\sigma})+
      \frac{\alpha}{2}(c_{j\downarrow}^\dag c_{j+1\uparrow}
      -c_{j+1\uparrow}^\dag c_{j\downarrow}-c_{j\uparrow}^\dag
      c_{j+1\downarrow}+c_{j+1\downarrow}^\dag
      c_{j\uparrow})\right]\;.
  \end{aligned}
  \label{eq:current}
\end{equation}

\begin{figure}[t]
  \center
  \includegraphics[width=1\textwidth]{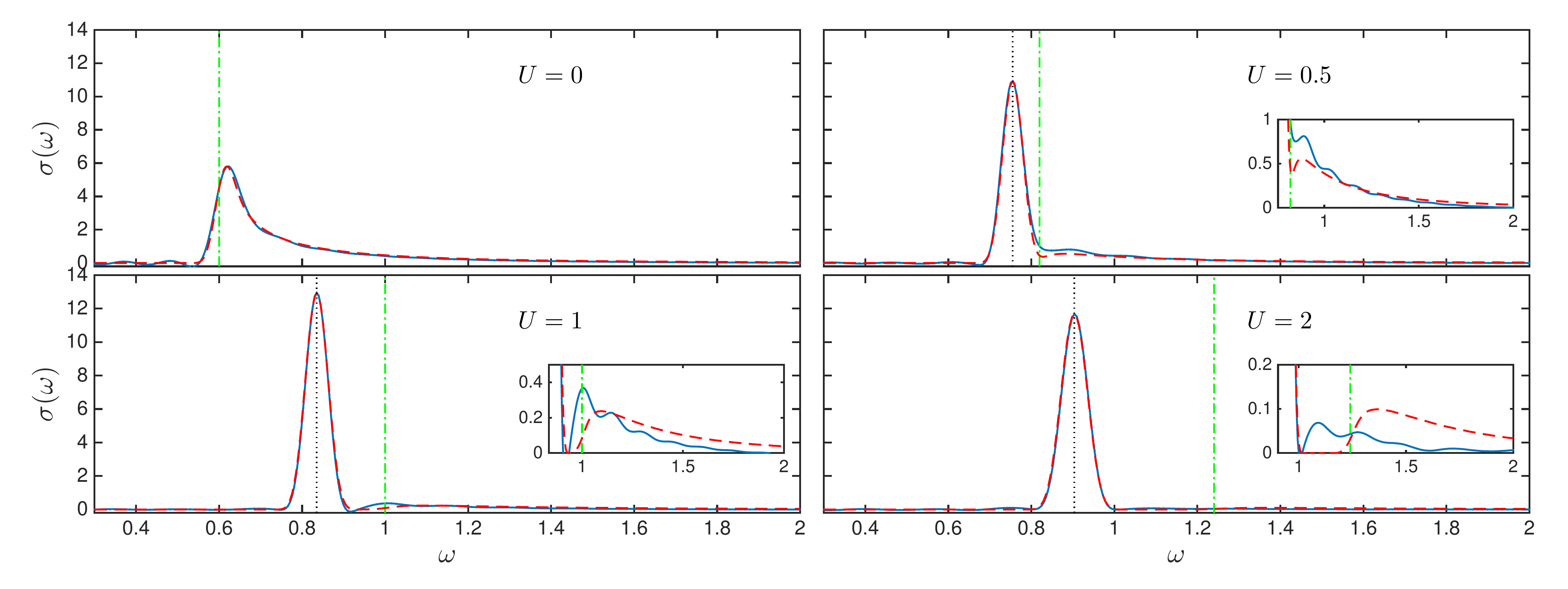}
  \caption{Optical conductivity $\sigma$ for $n=0.5$, $\alpha=1$,
    $B=0.3$, $L=200$, and different values of $U$.  The numerical
    result (solid blue line) is compared to the field-theoretical
    result (dashed red line) \cite{schuricht_spectral_2011}. The
    dot-dashed green line marks the SO gap $2 B^*$ 
    as obtained from figure \ref{fig:ldos}.  The dotted black line
    shows the extracted breather energy.  The insets zoom into the
    soliton-antisoliton continuum for better visibility.}.
  \label{fig:ocfb}
\end{figure}

The further calculation of $\sigma(\omega)$ is analogous to the
spectral densities and structure factors. We calculated the time
evolutions on a $L=200$ site system until $t_\mathrm{max}=42.5$. A
matrix dimension of $m=1200$ was used. We observed that the
entanglement grows slower with a $\hat J$ excitation than a single
particle excitation. Therefore, longer simulation times were feasible
at a comparable $\epsilon_\text{tot}$. Linear prediction was used up
to time 140.5. The time series was then multiplied with a window
function of Dolph-Chebyshev type.  The optical conductivity
$\sigma(\omega)$ of our microscopic model is presented in figure
\ref{fig:ocfb}~ \footnote{Apart from the imaginary part of
  $S_{\hat J \hat J}$, as in Eq.~\eqref{eq:oc}, one can also use the
  real part according to the Kramers-Kronig relations
  \cite{karrasch_transport_2014} in order to calculate
  $\sigma(\omega)$.  This was employed as a test; for $\omega > 0$ the
  two versions of $\sigma(\omega)$ are identical.}.

We are now ready to compare $\sigma(\omega)$ to the field theoretical
result $\sigma_{SLL}$ in a spiral LL \cite{schuricht_spectral_2011}.
The parameters for the field theoretical calculations are shown in
table~\ref{tab:par}.  Apart from $K_\sigma$, which was extracted from
the energy of the breather peak depicted by the dotted black lines in
figure~\ref{fig:ocfb}, all other parameters were obtained from
calculations independent of the optical conductivity.

The field theoretical result, $\sigma_{SLL}(\omega)$, was convoluted
with a Dolph-Chebyshev window, the same way as the numerical data, and
scaled such that the breather peak heights coincide, see
figure~\ref{fig:ocfb}. Its shape agrees very well with the numerical
result both in the interacting and in the noninteracting case. In the
latter there is no breather satisfying Eq.~\eqref{eq:breather},
instead the peak is given by the onset of the soliton-antisoliton
continuum at $\omega=2B$. This onset is moved to $\omega=2B^*$ in the
interacting case. The insets show the soliton-antisoliton continuum in
detail. We believe that the small-scale oscillations are artifacts
originating from the window function. At nonzero interactions, the
breather contribution at energies $\omega=\Delta_1 < 2 B^*$ emerges,
which in the field theory takes the form
$\sim\delta(\omega-\Delta_1)$. We note that we are always in the
parameter range $4/3 \le \kappa < 2$, where only a single breather
exists. Generally, we observe that the intensity of the field
theoretical result drops more slowly at high energies than in our
numerical calculations, which may originate from the existence of a
finite band width in our lattice model. To sum up, we conclude that
our simulations for the optical conductivity are in good agreement
with the field theoretical results.

\begin{figure}[t]
  \center
  \includegraphics[width=0.55\textwidth]{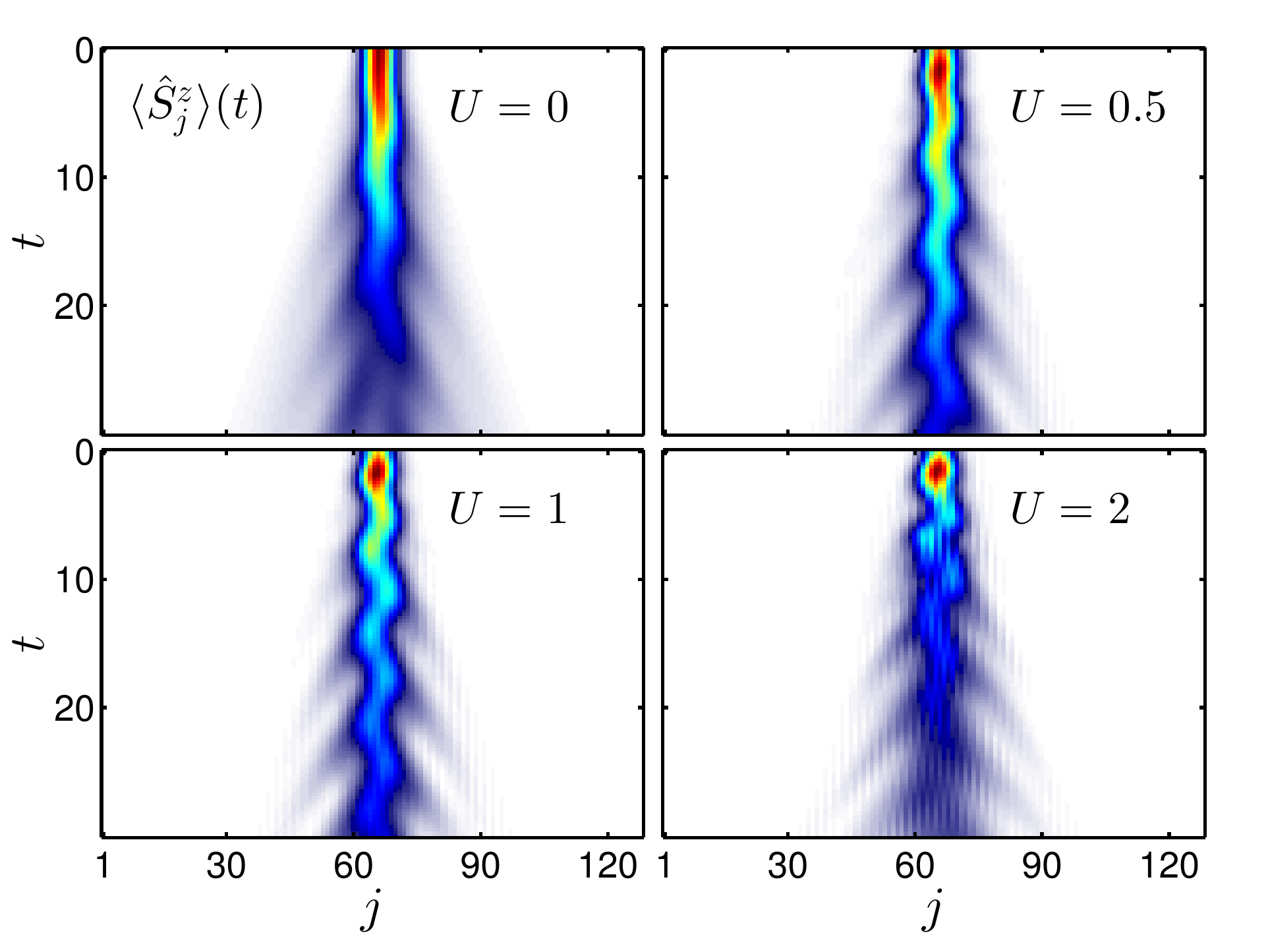}
  \caption{Time evolution of $\langle \hat S^z \rangle$ after an
    excitation \eqref{eq:gexc} for $n=0.5$, $\alpha=1$,
    $B=0.3, L=128,$ and different values of $U$.  }
  \label{fig:breather}
\end{figure}

\subsubsection{Time evolution of breather bound states.}
It is interesting, and potentially relevant for experiments, to
visualize the breather oscillations directly, by examining the time
evolution of the system after a local excitation from the ground
state.  A gaussian density excitation, centered around $k=0$, is
suitable for this task:
\begin{equation}
  \begin{aligned}
    & |\psi(t=0\rangle = g^\dag_{k=0} g_{k=0} |0 \rangle,  ~~~\text{where}\\
    & g_k=\sum_j e^{-\frac{(j-j_0)^2}{2\tilde \sigma^2}}
    e^{i(j-j_0)k}(c_{j\uparrow}^\dag+c_{j\downarrow}^\dag)\;.
  \end{aligned}
  \label{eq:gexc}
\end{equation}
We take $\tilde\sigma=4$, which corresponds to a width of $0.25$ in
$k$-space.  By employing an excitation in the eigenstate direction
$(c_{j\uparrow}^\dag+c_{j\downarrow}^\dag)$ with positive
$\hat S^x$-eigenvalue, we ensure that the excitation can act on both
branches of the dispersion simultaneously. The breather state itself
has a positive magnetization $\hat S^z$, which we use as the
observable in figure \ref{fig:breather}. The zigzag oscillations of
the breather state are clearly visible in the time evolution, with the
frequency of the oscillation corresponding to its energy.  The
oscillations at $U=0$ reflect the onset of the soliton-antisoliton
continuum at $\omega = 2 B$.  The breather energies in the optical
conductivity and for the direct excitation are in good agreement, see
table~\ref{tab:par}.  We observe that at $U=0.5$ and $U=1$ the
oscillations are longer lived than at $U=2$, which is a sign of the
onset of the Mott instability in the latter case.

\section{Conclusions and Outlook}
\label{sec4}

We have presented a detailed analysis of the static and dynamic
properties of strongly correlated quantum wires with Rashba SO
interaction and Zeeman field. We investigated a microscopic model,
with SO interaction, Zeeman field and tunable interactions of extended
Hubbard type, and calculated the static and dynamic properties by DMRG
and TEBD. We assessed the validity of the field-theoretical
description by comparing the results for the microscopic model to the
predictions for the corresponding low energy models, the helical and
spiral LL. In particular, we confirmed the enhancement of the SO gap
with increasing Coulomb interaction.  Furthermore, from the LL
parameters we determined the phase diagram of the system.  We found
that the parameter range (in filling, respectively SO interaction) of
the metallic 2F phase increases in the presence of interactions, and
that helical spin order and spin-dependent transport are preserved.
This means that interactions are a way to increase the SO gap without
disturbing the helical spin order (which a larger magnetic field would
do).  The interesting 2F phase thus becomes more accessible in
presence of interactions, which is very welcome in view of future
applications exploiting the helical spin order like spin-filters,
Cooper-pair splitters and Majorana
wires~\cite{oreg_helical_2010,lutchyn_majorana_2010,birkholz2008spin,streda_antisymmetric_2003,sato_cooper_2010}. The
main prediction of our work with respect to Majorana experiments is
the interaction enhancement of the SO gap.  In principle, this could
be detected by measuring the Land\'e $g$-factor in the nanowires. To
make a direct comparison, more information on the interaction strength
in semiconductor nanowires would be required, though. For very strong
interaction strengths, however, the 2F phase is suppressed in favor of
a Mott insulating phase in the commensurate case, characterized by the
opening of a charge gap.

Furthermore, we analysed characteristic breather bound states in the
optical conductivity $\sigma(\omega)$ as predicted for the spiral LL.
Using the extracted LL parameters, the optical conductivity was found
to be in good agreement with the field-theoretical results.  Finally,
we showed the presence of strong oscillatory behavior in the time
evolution of the bound states after an excitation, which can provide a
route for their experimental detection i.e. in cold atom systems.

While the present work focuses on the 2F phase, the LL theory predicts
interesting phases hosting fractional excitations if the chemical
potential is tuned below the SO
gap~\cite{frac1,meng_low_2014,frac2,frac3,frac4}. Using our methods, a
systematic study of these fractional phases in a microscopic model
could be addressed.

\section*{Acknowledgements}

We thank B. Braunecker, S. Grap and V. Meden for valuable
discussions. We acknowledge financial support from the Austrian
Science Fund (FWF), SFB ViCoM F4104, from Deutsche
Forschungsgemeinschaft (DFG) through ZUK 63, and SFB/TRR 21 from Nawi
Graz and from the National Science Foundation under Grant No. NSF
PHY11-25915.  MG acknowledges support by the Simons Foundation (Many
Electron Collaboration) and the Perimeter Institute.  Research at
Perimeter Institute is supported by the Government of Canada through
Industry Canada and by the Province of Ontario through the Ministry of
Economic Development \& Innovation.  This work is part of the D-ITP
consortium, an NWO program funded by the Dutch Ministry of Education,
Culture and Science (OCW).

\pagebreak
\appendix

\section{Linear prediction}
\label{linear_prediction}
The linear-prediction technique approximates future values of an
equally spaced time series $\{ y_i \}$ as a linear combination of past
values~\cite{white_spectral_2008,numerical_recipes_2007,Ganahl_Efficient_2014}
\begin{equation}
  x_l \approx \tilde x_l = -\sum_{j=1}^N a_j x_{l-j}\;.
\end{equation}
The coefficients $a_j$ are determined by minimizing a least squares
error.  Linear prediction efficiently extrapolates future data points
as a sum of damped exponentials, respectively Lorentzians after the
Fourier transformation, which is justified in many cases.  Due to the
large number of coefficients $a_j$, other functions can be represented
with sufficient accuracy as well.  We use $N_\text{max}/2$
coefficients, with $N_\text{max}$ being the number of steps in our
time evolution. The numerical effort of the prediction is negligible
compared to the time evolution itself.

\section{Spectra for Hubbard-type interactions}
\label{us0}
\begin{figure}[htb]
  \centering
  \includegraphics[width=0.66\textwidth]{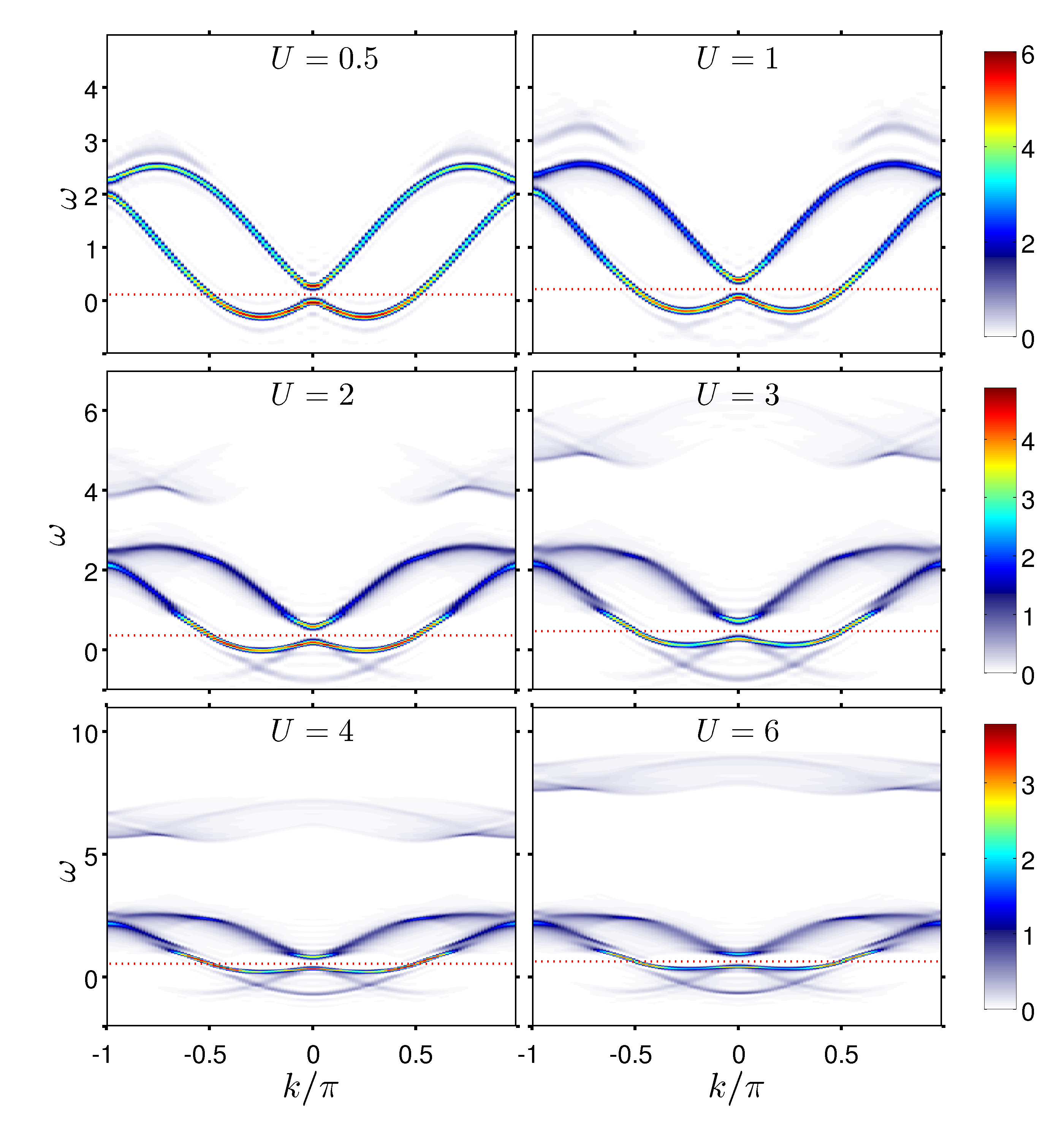}
  \caption{$S_{cc^\dag}(k,\omega)$ for $n=0.5$, $\alpha=1$, $B=0.1$,
    $L=128$ and several values of $U$ ($U'=0$). The dotted red line
    indicates the Fermi energy.  }
  \label{fig:specup0}
\end{figure}

In this appendix, we provide the spectral functions without
nearest-neighbor interaction $U'$, see figure \ref{fig:specup0},
whereas in the rest of the paper we have always taken $U'=U/2$ in
order to minimize backscattering.  Similar spectra, albeit for $B=0$,
have been obtained by QMC in \cite{goth_equivalence_2014}.  The
interaction effects in figure \ref{fig:specup0} are less pronounced
than in the previous results, due to the overall reduction of the
interaction.  In particular, the enhancement of the SO gap is reduced.
The two main branches of the spectral function preserve their general
shape and energy range for all values of the interaction.  As expected
\cite{giamarchi2004quantum} no Mott phase develops until $U=6$.  We
note that there is now a disjunct upper Hubbard band.

\section{Strong coupling region at incommensurate filling}
\label{cdw}
\begin{figure}[htb]
  \centerline{
    \subfloat[]{\includegraphics[width=0.407\textwidth]{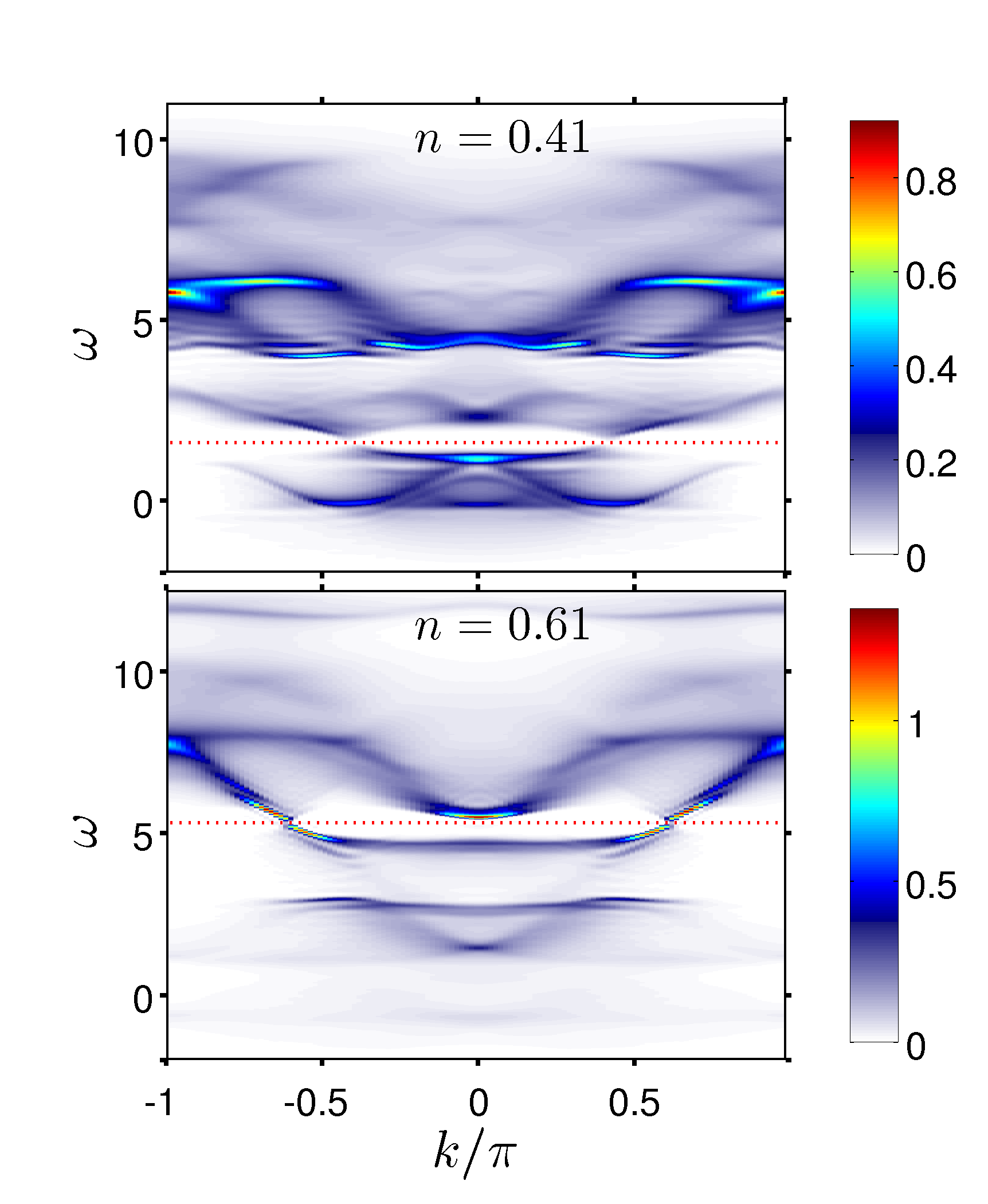}}
    \subfloat[]{\includegraphics[width=0.66\textwidth]{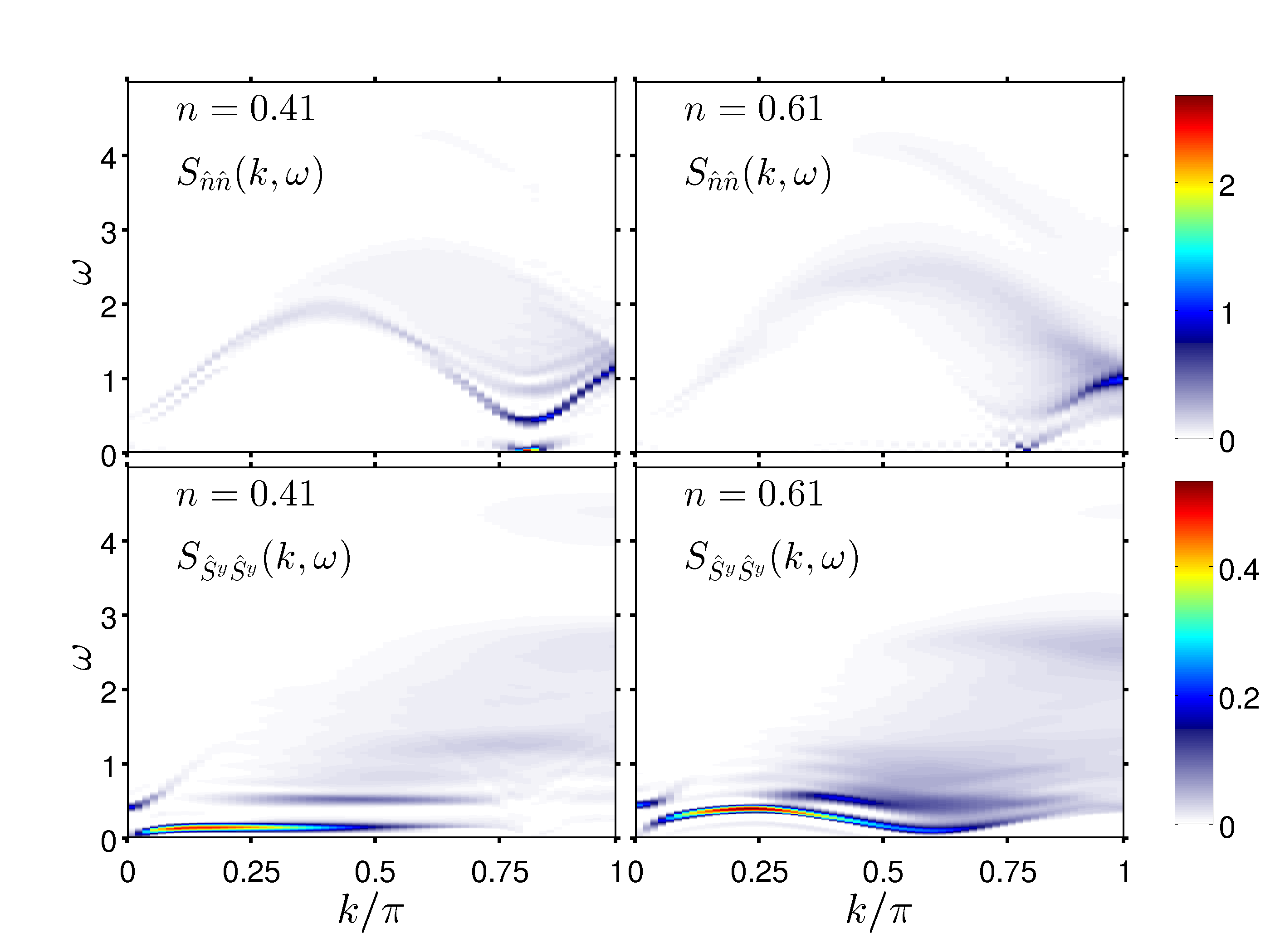}}}
  \caption{(a) Spectral functions and (b) charge and spin structure
    factors $S_{\hat n \hat n}(k,\omega)$ and
    $S_{\hat S^y \hat S^y}(k,\omega)$ for $n=0.41$ and $n=0.61$, with
    $\alpha=1$, $B=0.1$, $U=6$, $U'=U/2$ and $L=128$.  }
  \label{fig:spec_sc}
\end{figure}

In figure~\ref{fig:spec_sc} we provide the spectral functions and
structure factors for two different filling factors at a large value
of $U=6$ in the strong coupling region. For both $n=0.41$ and
$n=0.61$, the charge gap determined by Eq.~\eqref{eq:tpe}
vanishes, although the spectral functions shown in
figure~\ref{fig:spec_sc}~(a) exhibit a reduced spectral density around
the Fermi level.  The charge structure factor
$S_{\hat n \hat n}(k,\omega)$ displayed in
figure~\ref{fig:spec_sc}~(b) presents its largest contribution at
$k=2 k_\text F$ and $\omega=0$, indicating a tendency towards a charge
density wave phase, while the spin structure factor shows a
qualitatively similar behavior as for commensurate filling at smaller
interactions.

\FloatBarrier

\bibliography{literature}

\end{document}